%% file: McWhorterMultiChannel2023Feb.tex
\documentclass[10pt]{article}

\newcommand{\DocTitle}{ First-Order Statistical Framework for \\
Multi-Channel Passive Detection}
\newcommand{\ThanksText}{This work was supported in part
    by  the US Office of Naval Research (ONR) under contract
    N00014-21-1-2145, and by the US  Air Force Office of Scientific
    Research (AFOSR) under contract  FA 9550-14-C-0053.}

\title{ {\DocTitle} \thanks{\ThanksText} }
\author{Todd McWhorter \thanks{Brooks Canyon, LLC.   Deceased.}, 
Louis Scharf \thanks{Colorado State University, {\tt scharf@colostate.edu}}, and Margaret Cheney\thanks{Colorado State University, {\tt cheney@math.colostate.edu}}}

\usepackage[T1]{fontenc}          
\usepackage[scaled=0.92]{helvet}  

\usepackage{amsmath}              

\usepackage{newtxmath}

\usepackage{subfigure}
\usepackage{rotating}
\usepackage[dvipsnames]{xcolor}  
\usepackage{graphicx,psfrag}
\usepackage{realboxes}  
\usepackage{float}
\usepackage{array}
\usepackage{multirow}
\graphicspath{{./Figures/}}

\usepackage{fullpage}
\usepackage{afterpage}

\floatstyle{plain}
\restylefloat{table}

\usepackage{color}

\usepackage{cite}

\usepackage{xspace}
\input basedef2e.tex

\newcommand{\AZhat}{\widehat{\bfA}_Z} 
\newcommand{\AXhat}{\widehat{\bfA}_X} 
\newcommand{\AYhat}{\widehat{\bfA}_Y} 

\newcommand{\RZZ}{\bfR_{ZZ}}
\newcommand{\RZZhat}{\widehat{\bfR}_{ZZ}}

\newcommand{\SZZ}{\bfS_{ZZ}}
\newcommand{\SZZtil}{\widetilde{\bfS}_{ZZ}}

\newcommand{\QXX}{\bfQ_{\,XX}}
\newcommand{\QYY}{\bfQ_{\,YY}}
\newcommand{\QZZ}{\bfQ_{\,ZZ}}

\newcommand{\QEE}{{\bfQ_{\,EE}}}

\newcommand{\QEEinv}{{\bfQ^{-1}_{EE}}}

\newcommand{\cvsymb}{V}

\newcommand{\CVZcfar}{{\cvsymb_{Z,\text{CFAR}}}}

\newcommand{\SDE}{\calE}
\newcommand{\SDEtil}{\widetilde{\SDE}}

\newcommand{\cct}[1]{\vert c_{#1} \vert^{\,2}}

\newcommand\ReallyBigTstrut{\rule{0pt}{12.0ex}}    
\newcommand{\EndStrut}[1]{\rule[-#1]{0pt}{#1}}
\newcommand{\StartStrut}[1]{\rule[#1]{0pt}{#1}}

\usepackage[normalem]{ulem}     
\newcommand{\redout}{\bgroup\markoverwith{\textcolor{red}
                    {\rule[0.5ex]{10pt}{0.4pt} }} \ULon}

\begin{document}
\maketitle

\begin{abstract}
  \noindent In this paper we establish a general first-order
  statistical framework for  the detection of a common signal impinging on  spatially 
  distributed receivers.   
 We consider three types of channel models:  
  1) the propagation channel is completely known, 2) the propagation is known but channel gains are unknown, and
  3) the propagation channel is unknown.   For each problem, we address the cases of a) known noise variances,
  b) common but unknown noise variances, and c) different and unknown noise variances.   
    For all 9 cases, we establish generalized-likelihood-ratio (GLR) detectors, and show that each one  
  can be decomposed into two terms.  The
  first term is a weighted combination of the GLR detectors
  that arise from considering each channel separately. This 
  result is then modified by a fusion or cross-validation term,
  which expresses the level of confidence that the single-channel
  detectors have detected a common source.  Of particular note are the
  constant false-alarm rate (CFAR) detectors that allow for 
  scale-invariant detection in multiple channels  with different  
  noise powers. 
\end{abstract}

\tableofcontents

\section{Introduction}

In this paper we establish a general framework for the detection of a
signal that is common to two or more disparate measurement channels.  
This framework is based on a first-order linear 
model for the multivariate normal measurements in each channel.
An example of this scenario is the  use of passive, 
spatially-separated  arrays of sensors to detect a source radiating
acoustic or electromagnetic energy.  In this case, 
the question to be answered is whether or not the measurements 
contain a signal common to all sensor arrays, indicating the existence of a
radiating source.

This work addresses the case in which various quantities are unknown or uncertain.   
The uncertain quantities include not only the transmitted signal, but potentially also the precise position of the arrays, the environment through which the waves propagate, and  different noise levels on the various arrays.

The detection statistics herein are 
generalized likelihood ratios (GLRs), i.e. the statistics are a ratio of
likelihoods, each of which is separately maximized with respect to
unknown parameters in a measurement model.  The aim is to maximize the
output signal-to-noise ratio (SNR) of a multi-channel receiver. These GLRs
take many forms, depending upon which parameters are unknown. When
the various coefficients of our detectors are parameterized, for
example, by range, Doppler, geographic coordinates, etc., then the
detector statistics may be scanned through these parameters to
generate what might be called ``likelihood images''.

Within this paper we establish a general structure whereby the
composite multi-channel detector is a
weighted combination of the detectors specific to each channel,
and this combination is then modified by a fusion or cross-validation term.
The weights, which sum to one, may be interpreted as an {\it a priori}
confidence in each channel's detection statistic.
The cross-validation term is a measure of the correspondence between
the single channel detectors.  How one measures correspondence is a
function of what is known and unknown in the parameter space. 
 Many of these measures of correspondence are nonlinear; consequently
although the underlying
measurement model is a first-order model, 
the resulting GLR statistics are decidedly nonlinear functions of the measurements. 

\subsection{Relation to the literature}

In this section we describe the connections and differences of the
results in this paper to some existing work in the literature.  The first
observation is that the model and the framework established herein is
general and is not restricted to any particular physical model.  That
said, much of the relevant literature is concerned with the radar
problem described in Section~\ref{sec:motivation}.  Consequently, it
is this literature and problem to which we relate our work.
Within this literature it is assumed that the objective of the
measurement/processing system is to infer the presence and possibly
the location of source(s) of electromagnetic radiation. 

It is necessary to clarify those collection scenarios to which our
work is applicable.  First, we assume that the transmitted signal is
 unknown except for possibly its bandwidth.
Consequently the results herein are not, in their present form,
applicable to the multi-static active radar problem where the signal
is assumed known.  

A second class of problems consists of ``passive radar", scenarios
where a scene is illuminated by a source of opportunity, the
waveform of which is not assumed known. 
For many such scenarios, some of the receiver(s) (often called reference
arrays) always receive a scaled, delayed and noisy version of the
transmitted signal.  At the same time, other receivers (often called
surveillance arrays) are used to detect reflected radiation and
these arrays may or may not have a direct-path signal.   Despite the
fact that the signal is unknown, our hypotheses do \emph{ not} match this
scenario as the reference arrays always measure ``signal'' regardless
of the presence or absence of a target.  Examples of this detection
scenario include \cite{gogineni2017passive}, \cite{zhang2017direct}.
 If the collection geometry is such that \emph{no} channels receive
a version of the transmitted signal (direct- and/or multi-path)
when a target is not present, then
the results herein \emph{are} applicable.  See for example
\cite{Bialkowski11} and \cite{liu2014two} for examples of this type of
measurement system. 

Finally, the results herein
are applicable to the detection and localization of a source, e.g a
radar, which transmits an unknown signal.

There are a number of approaches to this passive source detection problem.  
For example, 
detectors based on first-order models, are derived in 
\cite{Weiss}, \cite{Hack} and \cite{VanKay12}.
Our treatment 
differs from this work in that we factor the likelihood into
sensor-specific and sensor-coupling terms, we treat the case of 
unknown noise powers at each sensor array, and we treat the case of an
unknown channel between source and sensor.

Approaches that assign a prior distribution to the common signal are
reported in \cite{Cochran} and \cite{Santamaria}, where the model may
be said to be a second-order statistical model, and in 
\cite{Howard}, where the marginalized measurement densities are not
characterized by second-order covariance. A comparison of these
approaches with the first-order GLRs of this paper is a function 
of SNR,
number of sensor elements, number of measurements, number of sources,
and what is known or unknown in the assumed parametric model for measurements. 
Mismatch between the assumed statistical model and the ``true'' model
can greatly affect performance for both  first- and
second-order detectors.

We also broadly categorize methods as 
estimation/localization or as detection/localization.  In the
estimation/localization category, estimates of unknown source parameters,
including its location, are found by maximizing an objective function
(usually a likelihood function of the data).  See for example
\cite{Weiss} and \cite{weiss2004direct}.  A difficulty with using an
estimator to infer the presence of a source is that an estimate of
source location is found even if the data consist only of noise, which
can result in high-variance estimates of the source location over time.
In addition, if multiple sources are present, they may not be detected
or localized since only one source location is estimated in these
methods.

The detection/localization methods use a detection statistic rather
than an estimator to infer the presence of a source.  A subset of this
category consists of systems that produce a \emph{single} detection
statistic, which if compared to a threshold, produces a binary (source
or no-source) decision.  
This method can be subject to the same
instability in the estimates of the source locations but this is
somewhat mitigated by the value of the detection statistic, which
indicates the ``confidence'' one might ascribe to the estimate.
Examples within this category include \cite{VanKay12}, \cite{Bialkowski11},

A second approach to this problem is to compute detection
statistics for a set of posited source locations and velocities etc.  
With this approach, it is possible to produce an image in which the 
value of the detection statistic at a location indicate the likelihood that a source
is present at that location.   
This method has
the advantage that it is possible to detect more than one source.
This is the approach used in this paper and is also
used in some sections of \cite{VanKay12}.

\subsection{Contributions of this paper}

A contribution of this paper is to show that the first-order,
multi-channel, detection problems have a common detector structure
where the composite (multi-channel) detector is equivalent to a
weighted sum of the per-channel detectors, which is then diminished by
a cross-validation term.  The cross-validation, or fusion, term is the
only quantity in the expression that uses the data from all channels
and it encapsulates all the multi-channel aspects of the problem.
This term can be interpreted as an indicator of the correspondence
among the multi-channel detectors and often has an intuitive
interpretation.  We show how this structure is maintained for a
variety of known or unknown noise and/or signal parameters.

This multi-channel detection framework provides a flexible basis for
designing fusion topologies.
As an example, one can determine the
formulas for a ``daisy-chained'' topology where each link in the chain
fuses the results of the previous channels and provides an
intermediate multi-channel result.  Another possible topology is a
tree structure. We show how the formulas
can be used to determine the information each channel must transmit to
be fully incorporated into a fused result.  The consequences of a
disabled channel or communication path can be considered.
The presence or absence of intermediate fusion centers can also be accommodated.

We  derive detectors for three different channel models and for three different assumptions for noise variances.   
 The resulting detectors have a necessary scaling
invariance property and are of particular interest for partially coherent and non-coherent channels.


\section{Motivation}
\label{sec:motivation}

The development of the detectors in this paper is 
not explicitly coupled to any particular physical model for
the measurement system. This is intentional as we believe these
results are applicable to a variety of detection problems.  Here we
briefly outline a representative problem for which this detection
framework is applicable.  

Assume a measurement $\bfx_{\,\ell}$ is a
sampled time series measured at \mbox{sensor $\ell$}.  Let the time interval of
the measurement be $T$.  The source is assumed to
emit a real-valued, bandlimited waveform $s(t)$ that is deterministic but unknown.
The corresponding baseband waveform we denote by $w(t)$.   
The Fourier series representation of this waveform we write as 
\begin{equation}
   s(t) \eqdef \text{Re}\left\{ e^{i2\pi f_c t} w(t) \right\}
    = \text{Re} \left\{ e^{i2\pi f_c t} \sum_{j=0}^{J-1} a_j
   e^{i\frac{2 \pi j \EndStrut{0.35ex}}{\StartStrut{0.55ex}T} t } \right\} .
\end{equation}
Each sensor, for example \mbox{sensor $\ell$}, receives a delayed,
scaled, and noisy version of this transmission, namely
\begin{equation}
 g_{\,\ell} \, s\left( t-t_{\,\ell}-\tau_{\,\ell}(t) \right) + n_{\,\ell}(t)
 = g_{\,\ell} \, \text{Re}\left\{ e^{i2\pi f_c (t-t_{\,\ell}-\tau_{\,\ell}(t)) } w(t-t_{\,\ell}-\tau_{\,\ell}(t)) \right\} + n_{\,\ell}(t) .
\end{equation}
Here $\tau_{\,\ell}(t)$ represents the (possibly time-varying) propagation
delay between the source and sensor.  It encapsulates the speed of
propagation in the medium and the time-varying relative
positions of the source and sensor. The constant $t_{\,\ell}$ represents any offset
between the clock of the sensor and a reference clock.
Here $g_{\,\ell} \in \C$ is a channel-gain term, which can include any
sensor gain and any attenuation losses due to propagation.

The received waveform is then complex demodulated.  On
\mbox{sensor $\ell$} we denote the resulting complex-valued baseband signal by
\begin{align}
  &x_{\,\ell}(t) =  g_{\,\ell}\, e^{-i 2\pi f_c(t_{\,\ell}+\tau_{\,\ell}(t))}\,
  w(t-t_{\,\ell}-\tau_{\,\ell}(t)) + u_{\,\ell}(t)
  \nonumber \\
  &\:= 
   g_{\,\ell} \,e^{-i 2\pi f_c(t_{\,\ell}+\tau_{\,\ell}(t))}\, \sum_{j=0}^{J-1} a_j
    e^{ i\frac{2 \pi j \EndStrut{0.35ex}}{\StartStrut{0.55ex}T} (t-t_{\,\ell}-\tau_{\,\ell}(t)) } +u_{\,\ell}(t).
\end{align}
Here $u_{\,\ell}(t)$ denotes the combination of the demodulated RF noise
($n_{\,\ell}(t)$) and any sensor noise.

It is assumed that this waveform is sampled at rate $1/T_s$, which is
at or above the Nyquist frequency.  The measurement interval $T=N_{\,\ell}
T_s$ consists of $N_{\,\ell}$ of
these samples, which in vector form is
\begin{align}
  \bfx_{\,\ell} &= \fourvec{
    x_{\,\ell}(0)}{x_{\,\ell}(T_s)}{\vdots}{x_{\,\ell}((N_{\,\ell}-1)T_s)}
  \in \C^{N_\ell} \nonumber \\
  &\eqdef g_{\,\ell} \, \left[ e^{-i2\pi f_c t_{\,\ell}}\,\bfV_{\,\ell}\,
    \bfD_J(t_{\,\ell}/T) \right] \bfa + \bfu_{\,\ell}
  \nonumber \\
  &\eqdef  g_{\,\ell} \bfH_{\,\ell} \,\bfa + \bfu_{\,\ell}
\label{eq:H_broadband}
\end{align}
where $\bfa \in \C^J$ is a vector of the signal amplitudes.  
The $nj$th element of $\bfV_{\,\ell} \in \C^{N_{\,\ell} \times J}$ is
\begin{equation}
  \left[ \bfV_{\,\ell} \right]_{nj} =  e^{-i2\pi f_c \tau_{\,\ell}(nT_s)} \,
     e^{i \frac{2\pi n j\EndStrut{.35ex}}{\StartStrut{.55ex}N_{\,\ell}} } \,
      e^{-i \frac{2 \pi j \EndStrut{.35ex}}{\StartStrut{.55ex}T} \tau_{\,\ell}(nT_s) }
\label{eq:Vell_def}
\end{equation}
and in general $\bfD_p(z) = \text{diag}\left\{ 1,\, e^{-i2\pi z},\, \cdots
\, e^{-i2\pi(p-1)z} \right\}$.  The model in (\ref{eq:H_broadband}) is
the structure assumed in this paper.  

The matrix $\bfH_{\,\ell}$ defined above is general for the problem under
consideration: the signal can be broadband and no approximations are
made with regard to the time-varying delay.   Now assume 
that the time-varying delay can be well approximated by a
first-order Taylor series
\begin{equation}
  \tau_{\,\ell}(t) \approx \tau_{\,\ell}(0) + \frac{\nu_\ell}{f_c}\, t.
\label{eq:tau_taylor_series}
\end{equation}
In addition assume that the signal is sufficiently narrowband so that,
for $j=1,2,\cdots J$,
\begin{equation}
  \exp\left(i \frac{2 \pi j \EndStrut{0.35ex}}{\StartStrut{0.55ex}T}
  \frac{\nu_\ell}{f_c}\,t\right) \approx 1
  \qquad \text{on} \, [\,0,\,T\,].
\label{eq:narrowband_assumption}
\end{equation}
When the approximations of (\ref{eq:tau_taylor_series}) and
(\ref{eq:narrowband_assumption}) are incorporated into
(\ref{eq:H_broadband}) we obtain the channel model
\begin{align}
  \bfH_{\,\ell} = 
   e^{-i 2\pi f_c (t_{\,\ell} + \tau_{\,\ell}(0))}
   \bfD_{N_{\,\ell}}(\nu_\ell
   T_s) \, \bfV \, \bfD_J\left(\frac{t_{\,\ell} +
     \tau_{\,\ell}(0)}{T}\right).
\label{eq:narrowband_H}
\end{align}
Here $\bfV$ is a \mbox{$J$-column} slice of a DFT matrix.  The
diagonal, modulation-matrix $\bfD_{N_{\,\ell}}(\nu_\ell T_s)$ models any Doppler
shift and $\bfD_J(\cdot)$ accounts for the clock offset and the
time-zero propagation delay. Note that, 
in this narrowband case, $\bfH_{\,\ell}^{\,\HT} \, \bfH_{\,\ell} = N_{\,\ell}\,
\bfI_{N_{\,\ell}}$, which simplifies the form of the detectors that
follow.


\section{General linear model}
\label{sec:general_model}

Consider a measurement system with $L$ channels, each of which is
excited by the same source.
When signal is present in the data we model a measurement on
a single \mbox{channel $\ell$} as
\begin{equation}
  \bfx_{\,\ell} = g_{\,\ell} \, \bfH_{\,\ell} \, \bfa + \bfu_{\,\ell} \in
  \C^{\,N_{\,\ell}}.
  \label{eq:channel_ell_data}
\end{equation}
The vector $\bfu_{\,\ell}\sim\calC
\calN_{\StartStrut{0.8ex}N_{\,\ell}}\!(\bfzero,\bfR_{\,\ell\ell})$ is additive
Gaussian noise,  which is assumed to be uncorrelated
across channels.  The noise covariance matrix is factored as
\begin{equation}
  \bfR_{\,\ell\ell} = \sigma_{\,\ell}^2 \, \bfSigma_{\,\ell\ell};  \quad \tr (
  \bfSigma_{\,\ell\ell} ) = N_{\,\ell} 
\end{equation}
where the normalized covariance $\bfSigma_{\,\ell\ell}$ is assumed known or
is estimated using signal-free auxiliary data.  Then, without loss
of generality, $\bfx_{\,\ell}$, $\bfu_{\,\ell}$ and $\bfH_{\,\ell}$  can be
replaced 
by their respective whitened versions: e.g.
\mbox{$\bfx_{\,\ell} \leftarrow \bfSigma_{\,\ell\ell}^{-1/2} \,\,\bfx_{\,\ell}$}.
Then $\bfSigma_{\,\ell\ell} = \sigma_\ell^{\,2} \bfI_{N_\ell}$.

The matrices $\bfH_\ell \in \C^{\, N_\ell \times \, J}$ 
represent the coupling of the signal amplitudes to the respective
measurement channels. There is a scaling
ambiguity between the gain $g_\ell$ and any signal gain
provided by $\bfH_\ell$.  This is resolved, without loss of generality, by 
requiring $\tr \left( \bfH_\ell^\HT \, \bfH_\ell \right) = J$.  It
follows that $g_\ell$ will encapsulate all the channel gain for
\mbox{channel $\ell$}.
Depending on the underlying physical model, the bases for
$\subspace{\bfH_\ell}$ may be completely specified, or
constrained to have orthonormal columns, or possibly only the
dimension $J$ of the subspaces is known.  These cases are treated,
respectively, in
Section~\ref{sec:known_channel},
Section~\ref{sec:Unknown_channel_gains} and
Section~\ref{sec:unknown_channels}.

Note that the signal amplitudes 
$\bfa \in \C^J$ are the \emph{same} for all channels, 
i.e. $\bfa$ is not indexed by $\ell$.
The amplitude vector $\bfa$ is considered unknown and is not
described by any
probabilistic or deterministic model (e.g. it is not constrained to a
finite set of symbols).  It follows that the signal component defines the
mean of the distribution of the data; hence the model is a first-order
statistical model.

The composite model for all channels is 
\begin{align}
  \bfz &= \fourvec{\bfx_1}{\bfx_2}{\vdots}{\bfx_{L}} =
  \fourvec{ g_1 \bfH_1}{g_2 \bfH_2}{\vdots}{g_{L} \bfH_{L}} \, \bfa + 
  \fourvec{\bfu_1}{\bfu_2}{\vdots}{\bfu_{L}}  \nonumber \\[1.3ex]
   &\eqdef
  \fourvec{ \bfF_1}{\bfF_2}{\vdots}{\bfF_{L}} \, \bfa + 
  \fourvec{\bfu_1}{\bfu_2}{\vdots}{\bfu_{L}} \nonumber \\
  &\eqdef \bfF \, \bfa + \bfu \in \C^{N_Z}.
  \label{eq:z_composite_model}
\end{align}
where $N_Z = \sum_{\ell=1}^{L} N_\ell$ is the length of $\bfz$.
Here $\bfF_\ell = g_\ell \bfH_\ell$ and $\bfF \in \C^{N_Z \times J}$
is the composite channel matrix.
The composite noise vector has distribution
\mbox{$\bfu \sim \calC\calN(\bfzero,
\text{blkdiag}\{ \,\sigma_\ell^2 \,\bfI_{N_\ell}\,\})$}.

\subsection{Measurements and notation}
\label{sec:measurements}

We assume that $M$ data-vectors $\bfz[m]$ are obtained and
that the vectors, as a group, either have signal present or they all consist of
noise.  The signal amplitudes, if non-zero, are assumed to be different
for each measurement vector. The additive noise vectors are modeled as
independent over the measurement index $m$, however the noise variance
is assumed
to have the same (known or unknown) value throughout the collection interval.
The channel is assumed to be static, i.e. $\bfF$ is constant
throughout the collection interval.

It is convenient to organize the totality of the measurements $\{ \bfz[m] \}$
into the $N_Z \times \, M$ matrix
\begin{equation}
  \bfZ = \left[ \, \bfz[1] \,\, \bfz[2] \,\, \cdots \, \, \bfz[M] \,  \right]  
    = \fourvec{\bfX_1}{\bfX_2}{\vdots}{\bfX_L} \in \C^{\,N_z \times \, M},
\label{eq:Zdef}
\end{equation}
which, when signal is present, has the model
\begin{align}
  \bfZ &=  \bfF \,\left[ \, \bfa[1] \; \bfa[2]\; \cdots \; \bfa[M] \,
    \right]
  + \left[ \, \bfu[1] \; \bfu[2]\; \cdots \; \bfu[M] \, \right]
  \nonumber \\
   &\eqdef  \bfF \,\bfA + \bfU.
\label{eq:Zmodeldef}
\end{align}
The $J \times \, M$ matrix
\begin{equation}
\bfA = [\, \bfa[1] \; \bfa[2] \;  \cdots \; \bfa[M]\, ]  \in
C^{\,J \times \, M}
\label{eq:Adef}
\end{equation}
consists of unknown signal amplitudes.

The detectors of this paper are functions of the sample covariance and
cross-covariance matrices denoted by
\begin{equation}
  \bfS = \frac{1}{M}\bfZ\,\bfZ^\HT =
  \left[ \begin{array}{cccc}
      \bfS_{\,11} & \bfS_{\,12} & \cdots & \bfS_{\,1L} \\
      \bfS_{\,21} & \bfS_{\,22} & \cdots & \bfS_{\,2L} \\
      \vdots & \vdots &  & \vdots \\
      \bfS_{\,L1} & \bfS_{\,L2} & \cdots & \bfS_{\,LL} 
          \end{array} \right].
\label{eq:sample_covariances}
\end{equation}
The whitened versions of these matrices, e.g.
\begin{equation}
\bfSwtil_{\,ij} = \bfR^{-1/2}_{\,ii} \, \bfS_{\,ij} \, \bfR^{-1/2}_{\,jj} =
\frac{\bfS_{\,ij}}{\sigma_i \,\sigma_j},
\label{eq:whitened_matrices}
\end{equation}
are distinguished by including '$\;\widetilde{\phantom{s}}\;$' on
the symbols.  The same notation is used to denote whitened subspaces and
channel-gains (e.g. $\gwtil_\ell \eqdef  g_\ell/\sigma_\ell$). Using the
representation in
(\ref{eq:z_composite_model}), we can define the whitened model for
composite \mbox{channel-$Z$}:
\begin{align}
  \bfZwtil &  \eqdef \RZZ^{-1/2} \, \bfX = 
  \fourvec{(g_1/\sigma_1)\,\bfH_1}{(g_2/\sigma_2)\,\bfH_2}
          {\vdots}{(g_{L}/\sigma_{L}) \,\bfH_{L}} \bfA 
          + \fourvec{\bfU_1/\sigma_1}{\bfU_2/\sigma_2}{\vdots}
          {\bfU_{L}/\sigma_{L}}  \nonumber \\[1.5ex]
          &\eqdef \fourvec{\gwtil_1 \, \bfH_1}
          {\gwtil_2 \, \bfH_2}{\vdots}
          {\gwtil_{L} \, \bfH_{L}} \bfA
          + \fourvec{\bfUwtil_1}{\bfUwtil_2}{\vdots}
          {\bfUwtil_{L}} 
          \eqdef \fourvec{\bfFwtil_1}{\bfFwtil_2}{\vdots}{\bfFwtil_L}
          \, \bfA + \bfUwtil \nonumber \\
          &\eqdef \bfFwtil\,  \bfA + \bfUwtil.
\label{eq:whitened_composite}
\end{align}
In due course these terms will be used to define GLR detectors.

\section{Hypotheses and generalized log-likelihoods}

The hypotheses to be tested are
\begin{align}
  &H_1: \bfZ = \bfF\, \bfA + \bfU; \nonumber \\
  &H_0: \bfZ = \bfU
\end{align}
where, for example, under $H_1$
\begin{equation}
  \bfZ \sim \calC\calN_{MN_Z} \left( \bfF\, \bfA,\, \bfI_M
\otimes \text{blkdiag}\, \{\,\sigma_\ell^2 \, \bfI_{N_\ell}\,\}
\right).
\end{equation}
A variety of GLR detectors can be derived based on the various combinations of
known and unknown model parameters.  Some of these detectors, for the
single channel, are well
described in the literature, including the matched-subspace detector (MSD)
\cite{Scharf94} and the constant-false-alarm rate (CFAR) MSD detector
\cite{Scharf91}. There are multi-channel
analogs to these detectors for certain combinations of known or unknown
parameters.  

The statistics herein are logarithms of the ratio of generalized
likelihood functions.  The fact that we use the logarithm of the ratio
is a convenience and does not affect detector performance (the
transformation function is monotonic). A generalized likelihood
function is defined herein as the
maximum value of a likelihood function over the domain of 
the unknown parameters in the model.  In other words, the unknown
quantities in the likelihood function are replaced by their
corresponding  maximum-likelihood (ML) estimates.  
Note that
these estimates are different under each hypothesis.

\subsection{Generalized likelihood function: $H_0$}

Under $H_0$, the Gaussian log-likelihood function is
\begin{align}
&\calL_0(\RZZ(0);\bfZ) \eqdef \ln \{ \,\ell_0(\RZZ(0);\bfZ)\, \}
\nonumber \\
&\quad = -M \,\text{logdet} \{ \, \RZZ(0)\, \} 
 	 -M\,\tr \left( \RZZ^{-1/2}(0) \, \SZZ\, \RZZ^{-1/2}(0)
\right) \nonumber \\
&\quad \eqdef -M \, \text{logdet} \{ \, \RZZ(0)\, \}  
-M\, \tr \left(  \, \SZZtil(0) \right)
\label{eq:L0_gen}
\end{align}
where $\ell_0(\RZZ(0);\bfZ)$ denotes the likelihood function
for $\RZZ(0)$ under hypothesis $H_0$.
We have dropped the term $\ln(1/\pi^{MN_Z})$ that is common to
likelihood under both hypotheses.  We use $\RZZ(0)$ to indicate
$\RZZ$ has unknown components to be estimated under $H_0$. The same
notation is used for the whitened
sample covariance matrix 
$$\SZZtil(0) \eqdef \RZZ^{-1/2}(0) \, \SZZ \,
\RZZ^{-1/2}(0). $$
Equation (\ref{eq:L0_gen}) is applicable for any noise covariance matrix
$\RZZ(0)$.  In this paper we  consider only the case where the noise
on \mbox{channel $\ell$} is uncorrelated with the noise on a different 
\mbox{channel $n$}. Under
these conditions, and using the definitions in Section~\ref{sec:measurements},
the log-likelihood function can be written as 
\begin{align}
&\calL_0(\RZZ(0);\bfZ) =  -M \sum_{\ell=1}^L \bigg\{
\text{logdet} \{ \, \bfR_{\,\ell\ell}(0)\, \} + \tr \! \left(  
\bfSwtil_{\,\ell\ell}(0) \right) \! \!  \bigg\}
\nonumber \\
&\qquad=  -M \sum_{\ell=1}^L \bigg\{ N_\ell \,  \ln 
\{ \, \sigma_\ell^2(0)\, \} + \frac{\tr \left(  
\bfS_{\,\ell\ell}\right)}{\sigma_\ell^2(0)}  \bigg\}
\nonumber \\
&\qquad = - M\sum_{\ell=1}^L \calL_0(\sigma^2_\ell(0); \bfX_\ell)
\label{eq:factored_L0}
\end{align}
This equation serves as a basis for the derivation of the GLR
detectors that follow.

\subsection{Generalized likelihood function: $H_1$}

Under the alternative $H_1$, the Gaussian log-likelihood function is
\begin{align}
&\calL_1(\RZZ(1),\bfA, \bfF ;\bfZ) \eqdef \ln \left( \ell_1(\RZZ(1),
  \bfA, \bfF;\bfZ) \right) \nonumber \\
&\quad = -M \, \text{logdet}\{\, \RZZ(1)\, \}  - \tr \left(  (\bfZ - \bfF \bfA)^\HT \,
  \RZZ^{-1}(1) \, (\bfZ -  \bfF \bfA) \right) \nonumber \\
&\quad = -M \, \text{logdet} \{ \, \RZZ(1)\, \}  - \tr \left(  (\bfZwtil(1) - \bfFwtil \bfA)^\HT \,
  (\bfZwtil(1) - \bfFwtil \bfA) \right) 
\label{eq:L1_gen}
\end{align}
where $\RZZ(1)$ denotes noise covariance to be estimated under $H_1$ and 
\begin{equation}
  \bfZwtil(1) \eqdef \RZZ^{-1/2}(1) \, \bfZ .
\end{equation}
We do not indicate the hypothesis for $\bfFwtil$  or $\bfA$
as these terms are present only when $H_1$ is in effect.
The ML estimate of the amplitudes $\bf A$, using all the channels, is
\begin{equation}
 \AZhat = (\bfFwtil^{\, \HT} \, \bfFwtil )^{-1} \bfFwtil^{\, \HT} \,
 \bfZwtil(1),
  \label{eq:Aest}
\end{equation}
which, when inserted into the log-likelihood function (\ref{eq:L1_gen}),
results in a generalized log-likelihood function
\begin{align}
  &\calL_1(\RZZ(1),\AZhat, \bfF; \bfZ)   \nonumber \\
  &\quad = -M\, \text{logdet} \{ \, \RZZ(1) \, \}  - \tr \left(
    \bfZwtil^{\,\HT}(1)  (\bfI - \bfP_{\,\bfFwtil})\,\bfZwtil(1) \right)
  \nonumber \\
  &\quad = -M \,\text{logdet} \{ \, \RZZ(1) \, \}  -M\, \tr \left(
    (\bfI - \bfP_{\,\bfFwtil})\,\SZZtil(1) \right).
  \label{eq:L1_Aest}
\end{align}
Here
\begin{equation}
  \bfP_{\,\bfFwtil} = \bfFwtil \, (\bfFwtil^{\,\HT} \, \bfFwtil)^{-1} \,
  \bfFwtil^{\,\HT}
 \label{eq:PF_def}
\end{equation}
is an orthogonal projection matrix with range $\subspace{\,\bfFwtil}$.

\subsection{Canonical Detector Structures}
\label{sec:general_detector}

This section describes a general form of the composite detector
structure in this paper.
Again assume that the inter-channel noise is uncorrelated and write the
log-likelihood in (\ref{eq:L1_Aest}) as
\begin{align}
  &\calL_1(\RZZ(1),\AZhat, \bfF; \bfZ)  
  = -M\, \sum_{\ell=1}^L \bigg\{\text{logdet} \{ \,
  \bfR_{\,\ell\ell}(1) \, \}
  +  \tr \left( \,
  \bfSwtil_{\,\ell\ell}(1) \, \right) \bigg\}
  + M\, \tr \left( \, \bfP_{\,\bfFwtil} \; \SZZtil(1) \, \right).
\label{eq:factored_L1}
\end{align}
Note that for a single channel the log-likelihood function is
\begin{align}
  \calL_1(\bfR_{\, \ell\ell},\bfA,\bfF_\ell; \bfX_{\,\ell}) 
  &= -M \, \text{logdet} \{ \, \bfR_{\,\ell\ell}(1) \, \}  - \tr \left(\,   (\bfX_{\,\ell} - \bfF_\ell\, \bfA)^{\,\HT} \,
  \bfR_{\,\ell\ell}^{-1}(1)\, (\bfX_{\,\ell}-   \bfF_\ell\, \bfA)\, \right) \nonumber \\
  &= -M \, \text{logdet} \{ \, \bfR_{\,\ell\ell}(1) \, \} 
   - \tr \left(	  (\bfXwtil_{\,\ell}(1) - \bfFwtil_\ell\, \bfA)^{\,\HT}   \,(\bfXwtil_{\,\ell}(1)- \bfFwtil_\ell\, \bfA) \, \right).
\end{align}
The ML estimate of the signal amplitudes using
\emph{only} the data from  \mbox{channel $\ell$}, namely
$ \bfAwhat_\ell = (\bfFwtil_\ell^{\, \HT} \, \bfFwtil_\ell )^{-1}
 \bfFwtil_\ell^{\, \HT} \, \bfXwtil_{\,\ell}(1)$, can be used to compress
 the per-channel log-likelihood
\begin{align}
  \calL_1(\bfR_{\,\ell\ell},\bfAwhat_\ell,\bfF_\ell; \bfX_{\,\ell})
  & = -M \, \text{logdet} \{ \, \bfR_{\,\ell\ell}(1) \, \}  - \tr
  \left(\bfXwtil_{\,\ell}(1) (\bfI - \bfP_{\,\bfFwtil_\ell}) \,
  \bfXwtil_{\,\ell}(1) \right) \nonumber \\
  & = -M  \text{logdet} \{ \, \bfR_{\,\ell\ell}(1) \, \} -
  M\tr \left(\,(\bfI - \bfP_{\,\bfH_\ell})\, \bfSwtil_{\,\ell\ell}(1) \,
  \right).
  \label{eq:compressed_L1_single_chan}
\end{align}
Here we have used that fact that $\bfP_{\bfFwtil_\ell} = \bfP_{\gwtil_\ell
  \bfH_\ell} = \bfP_{\bfH_\ell}$.
This expression allows (\ref{eq:factored_L1}) to be written as
\begin{align}
  \calL_1 & (\RZZ(1),\AZhat, \bfF; \bfZ)  
  = -M\, \sum_{\ell=1}^L \bigg \{ \, \text{logdet}\, \{ \,
 \bfR_{\,\ell\ell}(1) \, \} + \tr \left( \bfSwtil_{\,\ell\ell}(1) \right)
  \bigg\} 
 + M\, \tr \left( \, \bfP_{\,\bfFwtil} \, \SZZtil(1) \, \right)
 \nonumber \\
  &= -M\, \sum_{\ell=1}^L \bigg \{\, \text{logdet}\, \{ \,
 \bfR_{\,\ell\ell}(1) \, \} + \tr (\left(\bfI-\bfP_{\bfH_\ell}\right)
 \bfSwtil_{\,\ell\ell}(1)) \bigg \}  -M\left[
 \sum_{\ell=1}^L 
 \tr \left( \, \bfP_{\,\bfH_\ell} \, \bfSwtil_{\,\ell\ell}(1) \,
 \right)-
 \tr \left( \, \bfP_{\,\bfFwtil} \, \SZZtil(1) \, \right)
 \right] \nonumber \\
 &\eqdef M\sum_{\ell=1}^L \calL_1(\sigma_\ell^2(1),\bfAwhat_\ell,\bfF_\ell;
 \bfX_\ell) + M\Vwtil.
\label{eq:L1_with_cv_term}
\end{align}
It follows that a general form for the composite GLR detector is
\begin{align}
&   \Lambdawtil_Z = \frac{1}{M} \left(
  \calL_1(\RZZ(1),\AZhat,\bfF;\bfZ)
  - \calL_0(\RZZ(0);\bfZ) \right) \nonumber \\
  &= \sum_{\ell=1}^L \frac{1}{M} \left(
  \calL_1(\bfR_{\,\ell\ell}(1),\bfAwhat_\ell,\bfF_\ell;\bfX_\ell) -
  \calL_0(\bfR_{\,\ell\ell}(0);\bfX_\ell) \right) - \Vwtil  \nonumber \\
  &= \sum_{\ell=1}^L \Lambdawtil_\ell - \Vwtil.
\label{eq:general_composite_detector}
\end{align}
This is the general structure of the composite detectors in this
paper. The detector consists of a linear combination of the per-channel
detectors modified by a cross-validation term.  The cross-validation
term is the only component of this expression that uses the composite
data and it solely encapsulates the multi-channel aspects of the
problem.  As a general comment, the cross-validation term, which is 
\begin{align}
  \Vwtil &= \sum_{\ell=1}^L \tr \left( \bfP_{\bfH_\ell} \bfSwtil_\ell(1) \right) -
  \tr \left( \bfP_{\,\bfFwtil} \, \bfSwtil_{ZZ}(1) \right) \nonumber \\
  &=\sum_{\ell=1}^L \widehat{\text{SNR}}_\ell - \widehat{\text{SNR}}_Z ,
\label{eq:general_cross_validation}
\end{align}
can be interpreted as a difference in estimated
signal-to-noise-ratios.  The cross-validation term has additional,
insightful, forms that are discussed subsequently.

The per-channel detectors in (\ref{eq:general_composite_detector})
can be written as
\begin{align}
  \Lambdawtil_\ell &= \ln \left( \frac{\abs{\,\bfR_{\,\ell\ell}(0)\,}}
             {\abs{\,\bfR_{\,\ell\ell}(1)\,}} \right) +
             \tr \left( \, \bfSwtil_{\,\ell\ell}(0) \right) - 
             \tr \left( \,(\bfI-\bfP_{\,\bfH_\ell} )\,
             \bfSwtil_{\,\ell\ell}(1) \right),
\label{eq:general_per_channel}
\end{align}
which, when $\bfR_{\,\ell\ell} = \sigma_\ell^{\,2} \, \bfI$, reduces
to
\begin{align}
  \Lambdawtil_\ell &= N_\ell \ln \left( \frac{\sigma_\ell^{\,2}(0)\,}
             {\sigma_\ell^{\,2}(1)} \right) +
             \frac{\tr ( \, \bfS_{\,\ell\ell})}{\sigma_\ell^{\,2}(0)} - 
             \frac{\tr \left( \, (\bfI-\bfP_{\bfH_\ell})\bfS_{\,\ell\ell} \right)}
                  {\sigma_\ell^{\,2}(1)}.
                  \label{eq:general_per_channel_wgn}
\end{align}


\section{Detectors: Known Channel $\bfF$}
\label{sec:known_channel}

In this section we derive estimators and detectors assuming that the
channel gains and channel matrices are known,
i.e. the composite channel matrix $\bfF$ is completely
specified.  We derive detectors for three different noise models.

Two of the resulting GLR detectors can be considered equivalent to
detectors for a
single distributed channel.  The first instance is the ``clairvoyant''
or idealized case where the only
unknown parameters in the hypotheses are the signal amplitudes $\bfA$.
The second instance is the constant-false-alarm-rate (CFAR) detector
that arises when the noise variances on each channel are unknown but
constrained to be the same.   It is this assumption that makes the
model equivalent to a single channel case.
The third noise model assumes noise variances in each channel are unknown and different.
This assumption makes the problem a mulit-channel detection problem.   

It will be shown that these detectors can be expressed as a weighted 
combination of the detectors obtained from each channel and this sum
is then modified by a cross-validation term.  Under the conditions of
this section, each channel can obtain an independent unbiased estimate
of the mode amplitudes using only its local data.  We note that the
cross-validation terms
in these cases are functions of the Euclidean distance between
each channel's estimate of the mode amplitudes (this
difference is whitened).


\subsection{Known $\bfF$: known $\bfR_{\,\ell\ell}$}
\label{sec:KnownF_and_R}

In this section we assume that the inter-channel noise is uncorrelated
and  that  the per-channel covariance matrices $\{\,\bfR_{\,\ell\ell} \}$
are known or are estimated using signal-free auxiliary (training)
data.
Since the noise covariances are completely specified, they need not be
estimated under either hypothesis and
$\bfR_{\,\ell\ell}(0) = \bfR_{\,\ell\ell}(1) = \bfR_{\,\ell\ell}$ and
$\bfSwtil_{\,\ell\ell}(0)=\bfSwtil_{\,\ell\ell}(1)=\bfSwtil_{\,\ell\ell}$.
It follows that the per-channel detectors of
(\ref{eq:general_per_channel}) are
\begin{equation}
  \Lambdawtil_\ell = \tr\left(\, \bfP_{\bfH_\ell} \, \bfSwtil_{\,\ell\ell}
  \,\right),
\end{equation}
and, using the results of Section~\ref{sec:general_detector},  a composite
GLR detector is
\begin{align}
  \Lambdawtil_Z &=  \sum_{\ell=1}^L \frac{1}{L} 
  \tr\left(\, \bfP_{\bfH_\ell} \, \bfSwtil_{\,\ell\ell} \right)  - \frac{1}{L}\left(
   \sum_{\ell=1}^L \tr \left(\, \bfP_{\bfH_\ell} \, \bfSwtil_{\,\ell\ell} \right) 
    -    \tr \left( \bfP_{\,\bfFwtil} \, \bfSwtil_{ZZ} \right)  \right)
  \label{eq:clairvoyant_form_1} \\
  &\eqdef \sum_{\ell=1}^L \alpha_\ell \,
  \Lambdawtil_\ell- \Vwtil_Z.
\label{eq:clairvoyant_form_1_sc}
\end{align}
We have included a scale factor, $\alpha_\ell = 1/L$,  that does not
affect detector
performance but brings the resulting detector into the canonical
form of section \ref{sec:general_detector}.  Note that $\sum \alpha_\ell = 1$, which is a characteristic of
our canonical form.

The identity in Appendix~\ref{sec:MultiChan_decomp} 
is used to derive  a different form for the cross-validation term.
For example consider a two two-channel case. Let
\begin{align}
  \bfAwhat_{\,\ell} = \left(\bfFwtil_{\,\ell}^{\,\HT} \, \bfFwtil_{\,\ell}\right)^{-1} \,
  \bfFwtil_{\,\ell}^{\,\HT} \, \bfX_{\,\ell} 
  \quad \sim  \calC\calN\left (\bfA,
   \left(\bfFwtil_{\,\ell}^{\,\HT} \, \bfFwtil_{\,\ell}\right)^{-1} \right) 
 \quad  \sim  \calC\calN\left(\bfA, \bfQwtil_{\,\ell\ell}\right)
\end{align}
be a per-channel estimator of the signal amplitudes.  Then 
it can be shown that the
cross-validation term can be written as
\begin{align}
  \Vwtil_Z &= \frac{1}{M} \tr \left(\bfAwhat_1 - \bfAwhat_2 \right)^{\,\HT} \,
  \bfQ_{EE}^{-1} \, \left(\bfAwhat_1 - \bfAwhat_2 \right) \nonumber \\
  &=  \tr \left( \bfQwtil_{EE}^{-1} \,
  \left(\bfAwhat_1 - \bfAwhat_2 \right)\,
  \left(\bfAwhat_1 - \bfAwhat_2\right)^{\,\HT} /M \right) \nonumber \\
  &\eqdef \tr \left( \bfQwtil^{-1}_{EE} \, \bfSwtil_{EE} \right)
\label{eq:cv_two_channel_all_known}
\end{align}
where
\begin{equation}
  \QEE = E \left\{\, \left(\bfAwhat_1 - \bfAwhat_2\right)\,
   \left(\bfAwhat_1 - \bfAwhat_2\right)^{\,\HT}\, \right\} 
   = \bfQwtil_{11} +   \bfQwtil_{22}
\label{eq:QEEdef_two_channel}
\end{equation}
is the covariance matrix of the difference in the amplitude estimates.
The cross validation term in (\ref{eq:cv_two_channel_all_known})
is a non-negative function of the difference in the amplitudes
estimates from each channel.

It is clear from (\ref{eq:clairvoyant_form_1}) that the
composite detector can also be written as
\begin{equation}
\Lambdawtil_Z = \frac{1}{L} \tr \left( \bfP_{\,\bfFwtil}\, \bfSwtil_{ZZ} \right).
\label{eq:clairvoyant_form_comp}
\end{equation}
This expression 
is derived in \cite{Scharf17multipulse},
although there the detector is not expanded into our
canonical form since only a single channel was being considered.  This reflects the fact that the composite detector, under these
conditions, is in effect a single-channel detector with
distributed data.

The detector structure in (\ref{eq:clairvoyant_form_1_sc}) 
is illuminating and is an instance of the canonical form
for the detectors derived in this paper. The composite detector
can be expressed as a weighted combination of GLR detectors for each
channel, which is then diminished by a fusion or cross-validation
term. We wish to emphasize that the per-channel detectors (say
$\Lambdawtil_\ell$) use only the data on the indicated
channel and are the
GLR detectors one would derive for a single channel.  It is only the
cross-validation term that uses the data from all channels.
The cross-validation term  
expresses the confidence in the individual-channel detector values: when
$\Vwtil$ is large, the confidence in the individual-channel  
detector values is small.
When this difference increases, the cross-validation term increases
and the composite detector output is reduced.   This is intuitive: a
large difference in the amplitude estimates should, and does, reduce
the likelihood that each channel is excited by the same amplitudes.
The detectors that follow have a similar structure although the 
constituents in the structure are different depending on which
parameters are assumed known.  Equations
(\ref{eq:clairvoyant_form_1_sc}) - (\ref{eq:QEEdef_two_channel}) 
characterize the detector of this section.  The nine panels in
Table~\ref{tbl:GLMdetectors} may be labeled as $P_{ij}; \, i,j=1,2,3$.
Then this result is summarized in panel $P_{11}$ with $\alpha_\ell$ and
$\Lambda_\ell$ defined in the column label and the channel matrices
defined in the row label.

\subsection{Known $\bfF$: common but unknown variance}
\label{sec:idealized_CFAR}

In this section the noise variance is constrained to be
equal on all channels but is considered to be unknown,
i.e. $\RZZ =\sigma^{\,2}\,\bfI_{N_Z}$.  The likelihood functions in
(\ref{eq:L0_gen}) and (\ref{eq:L1_Aest}) are maximized
with respect to the noise variance when
\begin{equation}
  \sigmawhat^{\,2}(0) = \frac{1}{N_Z} \tr\left(\SZZ\right) ; \quad
  \sigmawhat^{\,2}(1) = \frac{1}{N_Z} \tr \left(\left(\bfI-\bfP_\bfF\right)\SZZ \right).
\end{equation}
Using these estimates to compress the log-likelihood functions of 
(\ref{eq:L0_gen}) and (\ref{eq:L1_Aest}) and transforming their
difference with a monotonic function results in a composite
detector
\begin{equation}
  \Lambda_{Z,\text{CFAR}} = \frac{ \tr \left( \bfP_{\bfF}\, \SZZ
      \right)}{\tr \left( \SZZ \right) } = \frac{L}{\tr(\SZZ)}
  \Lambdawtil_Z \;\bigg\vert_{\sigma_\ell^{\,2}=1}.
  \label{eq:Lambda_Z_CFAR}
\end{equation}
Although derived as a multi-channel detector, this result is
essentially a single channel detector \cite{Scharf91} with distributed
data. It is considered a constant false alarm rate (CFAR) detector as it is
invariant to a scaling of the \emph{composite} data $\bfZ$. We
indicate this property by adding "CFAR" as subscripts to the detector symbols.  It
follows from (\ref{eq:Lambda_Z_CFAR}) that
\begin{align} 
  \Lambda_{Z,\text{CFAR}} &=
  \sum_{\ell=1}^L \frac{1}{\tr(\SZZ)} 
  \tr(\bfP_{\bfH_\ell} \, \bfS_{\,\ell\ell})  -
    \frac{L}{\tr(\SZZ)} \Vwtil_Z
  \;\bigg\vert_{\sigma_\ell^{\,2}=1} \nonumber \\
  &= 
  \sum_{\ell=1}^L \frac{\tr(\bfS_{\,\ell\ell})}{\tr(\SZZ)} \left( 
  \frac{\tr(\bfP_{\bfH_\ell} \,
    \bfS_{\,\ell\ell})}{\tr(\bfS_{\,\ell\ell})} \right) -
  V_{Z,\text{CFAR}}
    \nonumber \\
 &\eqdef
  \sum_{\ell=1}^{L} \alpha_\ell
  \Lambda_{\,\ell,\text{CFAR}}  - V_{Z,\text{CFAR}}
  \label{eq:Lambda_Z_CFAR_form2}
\end{align}
where now $\alpha_\ell = \tr(\bfS_{\,\ell\ell})/\tr(\SZZ)$ 
(note $\sum \alpha_\ell = 1$).  A per-channel detector under the
conditions of this section
\begin{equation}
  \Lambda_{\,\ell,\text{CFAR}} = \frac{ \tr(\bfP_{\bfH_\ell}
    \bfS_{\,\ell\ell})}{\tr ( \bfS_{\,\ell\ell})}
\label{eq:Lambda_ell_cfar}
\end{equation}
is invariant to unique scaling of its data and is thus
locally CFAR, which is indicated by its subscript.
It is evident that the composite CFAR detector has a structure that is
conceptually identical to that in (\ref{eq:clairvoyant_form_1_sc}).
That is, the composite detector is a weighted combination of the
individual-channel
detectors modified by a cross-validation term. But now, the 
detectors are CFAR and they are combined through a data-determined
weighted combination rather than the fixed 
averaging used in (\ref{eq:clairvoyant_form_1_sc}).  The cross-validation term
is invariant to a scaling of $\bfZ$ (as is indicated by the subscript).
Equations (\ref{eq:Lambda_Z_CFAR_form2}) - (\ref{eq:Lambda_ell_cfar})
describe the detector of this section.  These results are summarized
in panel $P_{12}$ in Table~\ref{tbl:GLMdetectors}.

\subsection{Known $\bfF$: noise variances different and unknown}
\label{sec:knownFZ_unknown_different_noise}

Consider the case where the data are governed by hypotheses $H_0$.
Using (\ref{eq:factored_L0}), the ML estimate for the noise variance
on \mbox{channel $\ell$} is found to be 
\begin{equation}
  \sigmawhat_\ell^{\,2}(0) = \frac{1}{MN_\ell} \tr
  \left( \, \bfX_\ell\, \bfX_\ell^\HT \, \right) =
   \frac{1}{N_\ell} \tr \left(\,  \bfS_{\,\ell\ell} \, \right).
\end{equation}
This result can be used to compress the log-likelihood function
in (\ref{eq:factored_L0})
\begin{align}
  \calL_0(\RZZhat(0); \bfZ) &=  \sum_{\ell=1}^{L}
  \calL_0(\sigmawhat_\ell^{\,2}(0); \bfX_\ell) \nonumber \\
  &=  -M \sum_{\ell=1}^{L} N_\ell \ln \{ \tr \left( \,\bfS_{\,\ell\ell}
  \, \right) \}.
\end{align}
We have ignored various additive and multiplicative
constants whose inclusion or exclusion does not affect the detector
structure.

Rewriting (\ref{eq:L1_with_cv_term}), the log-likelihood function
under $H_1$ is
\begin{align}
  &\calL_1(\RZZ(1),\AZhat, \bfF; \bfZ)   \nonumber \\
  &\quad = -M\, \sum_{\ell=1}^L \bigg \{ N_\ell \, \ln \left\{ \,
  \sigma_\ell^{\,2}(1) \, \right\} + \frac{1}{\sigma_\ell^{\,2}}
  \tr (\left(\bfI-\bfP_{\bfH_\ell}\right) \bfS_{\,\ell\ell})  \bigg \}
 + M\Vwtil.
\label{eq:L1_with_cv_redux}
\end{align}
The procedure for finding the exact ML-estimates of the
noise variances involves
solving a coupled set of non-linear equations.  These equations can be solved
iteratively, 
however we find that approximate, per-channel, estimates suffice.
This choice also allows the resulting composite (multi-channel)
detector to have our
canonical form and desirable invariance properties.
If we let the cross-validation term be zero, we can use 
(\ref{eq:L1_with_cv_redux}) to \emph{locally} estimate the noise
variance
\begin{align}
  \sigmawhat_\ell^{\,2}(\Vwtil= 0) &= \frac{1}{N_\ell} \tr\left(\,
  (\bfI-\bfP_{\bfH_\ell})\, \bfS_{\,\ell\ell} \, \right).
\label{eq:per_channel_noise_est}
\end{align}
It follows that compressing  (\ref{eq:general_composite_detector}) and
(\ref{eq:general_per_channel_wgn}) with these estimates 
(and dividing by $N_Z$) gives
\begin{align}
  &\Lambda_{\{X_\ell \},\text{CFAR}} 
 = \sum_{\ell=1}^{L}
   \frac{N_\ell}{N_Z}  \ln \left\{  
  \frac{\tr\left(\bfS_{\,\ell\ell}\right)}
       {\tr\left((\bfI-\bfP_{\bfH_\ell})\bfS_{\,\ell\ell}\right)}
       \! \right\}  
   - \frac{1}{N_Z}
        \Vwtil\bigg\vert_{\sigma_\ell^{\,2}=\sigmawhat_\ell^{\,2}(1)}
        \nonumber \\
   &\quad \eqdef 
       \sum_{\ell=1}^{L} \alpha_\ell \Lambda^{(2)}_{\ell,\text{CFAR}}  - 
       V_{\{X_\ell\},\text{CFAR}}
\label{eq:unknown_diff_noise_det}
\end{align}
where $\alpha_\ell = N_\ell/N_Z$.
The composite detector has our canonical structure of a weighted
combination of detectors for
each channel, which is then diminished by a cross-validation term.
The subscript
notation on these detectors and terms indicates the scale invariances
of the term.  In particular $\Lambda_{\{X_\ell\},\text{CFAR}}$
indicates that
the composite detector is invariant to \emph{different} scalings of each
channel's data. 
This type of scaling-invariance is an essential property
for any realistic detector derived under the assumptions of this section.
The superscript '$(2)$' is used to indicate that the  per-channel 
detectors, e.g.
\begin{equation}
  \Lambda^{(2)}_{\ell,\text{CFAR}} = 
  \ln \left( \frac{\tr(\bfS_{\,\ell\ell})}{\tr( (\bfI -
    \bfP_{\bfH_\ell})\,\bfS_{\,\ell\ell})} \right) 
\label{eq:Lambda2_def}
\end{equation}
are different from the CFAR detectors in (\ref{eq:Lambda_ell_cfar}).
However, their distributions are similar.  The random variable in
(\ref{eq:Lambda_ell_cfar}) is beta distributed (centrally under $H_0$
and non-centrally under $H_1$).  Denote its density function by
$\text{Beta}(x)$.  Then the random variable in (\ref{eq:Lambda2_def}) will
have a density function $f(x) = e^{-x} \text{Beta}(1-e^{-x})$.
Equations (\ref{eq:unknown_diff_noise_det}) - (\ref{eq:Lambda2_def})
describe the detector under the conditions of this section, which is
summarized in panel $P_{13}$ in Table~\ref{tbl:GLMdetectors}.

\section{Unknown channel gains $\{ g_\ell \}$}
\label{sec:Unknown_channel_gains}

Within this section we assume that the channel gains are unknown.
In addition it is assumed that the channel matrices have the property
$\bfH_\ell^\HT \, \bfH_\ell = \bfI_J$. 
This may at first seem an
overly restrictive condition, but several problems, including those
described in Section~\ref{sec:motivation}, have channel matrices with
this property.
The resulting detectors will have the structure we
have discussed earlier.  But now, the cross-validation term is a
function of the \emph{coherence} between each channel's estimate of the
signal amplitudes, rather than a function of the
\emph{Euclidean distance} between these estimates.

\subsection{Unknown $\{ g_\ell \}$: known noise variances}

When we assume that the noise variances are known, the remaining
unknowns (the channel gains), are only present when $H_1$ is in
effect.  Consequently the GLR detector for these conditions can be obtained
by maximizing the clairvoyant detector in
(\ref{eq:clairvoyant_form_1}) with respect to the channel gains.

Under the conditions of this section we note that
\begin{align}
  \bfFwtil^{\,\HT} \, \bfFwtil &=
  \sum_{\ell=1}^{L} \magsq{ \, \gwtil_\ell}\,
  \bfH_\ell^{\,\HT} \, \bfH_\ell = 
  \sum_{\ell=1}^{L} \magsq{ \, \gwtil_\ell} \,
  \bfI_J \nonumber \\
  &= \left(\, \bfgwtil^{\,\,\HT} \, \bfgwtil\,\right) \, \bfI_J
  \label{eq:gtil_def}
\end{align}
where
\begin{equation}
  \bfgwtil^{\,\,\HT} \eqdef \left[ \,
    \frac{g_1^{\;\mbox{\normalsize$\ast$}}}{\sigma_1} \:\:
   \frac{g_2^{\;\mbox{\normalsize$\ast$}}}{\sigma_2} \:\: \cdots \: \:
   \frac{g_{L}^{\;\mbox{\normalsize$\ast$}}}{\sigma_{L}}\, \right].
\end{equation}
It follows that (\ref{eq:clairvoyant_form_1}) can be written as
\begin{align}
  \Lambdawtil_Z(\,\bfgwtil\,) &=
  \frac{1}{ \bfgwtil^{\,\,\HT} \, \bfgwtil}\:
  \tr \left( \frac{1}{ML}(\bfFwtil^{\,\HT} \, \widetilde{\bfZ} ) 
   \,(\bfFwtil^{\,\HT} \, \widetilde{\bfZ} )^{\HT} \right) \nonumber \\
   &= \frac{1}{ \bfgwtil^{\,\,\HT} \, \bfgwtil}\:
   \sum_{i=1}^{L} \sum_{j=1}^{L}\frac{1}{ML}
   \gwtil_i^{\,\mbox{\normalsize$\ast$}} \;\gwtil_j\;
   \tr \left( \bfH_i^{\,\HT} \, \bfXwtil_i \, \bfXwtil_j^{\,\HT} \, \bfH_j\right)  
    \nonumber \\
    &\eqdef \frac{\bfgwtil^{\,\HT} \,\bfMwtil_Z \,\, \bfgwtil}{
    \bfgwtil^{\,\,\HT} \,\bfgwtil}.
\label{eq:general_Rayleigh_quotient}
\end{align}
This Rayleigh quotient form, for the special case of equal noise
variances, was presented in \cite{VanKay12}.  
Recall that a single channel detector, for known noise variances, is
\begin{equation}
  \Lambdawtil_\ell = \frac{1}{M}
  \frac{ \tr \left( \bfH_\ell^{\,\HT} \, \bfX_\ell \, \bfX_\ell^{\,\HT} \,
    \bfH_\ell\right) }{\sigma_\ell^{\,2}} = \frac{\tr \left(
      \bfP_{\bfH_\ell} \,
    \bfS_{\,\ell\ell}\right) }{\sigma_{\ell}{\,^2}}.
\end{equation}
In Appendix~\ref{sec:unknown_gains_appendix}, 
  with all $\alpha_\ell = 1/L$, we demonstrate that 
(\ref{eq:general_Rayleigh_quotient}) can be written as
\begin{align}
  \Lambdawtil_Z &= \sum_{\ell=1}^L \alpha_\ell \Lambdawtil_\ell -
  \frac{ \bfgwtil^{\,\HT}\, \bfTwtil_Z \, \bfgwtil}{\bfgwtil^{\,\HT}
    \, \bfgwtil}
  \nonumber \\
  &\eqdef
  \sum_{\ell=1}^L \alpha_\ell \Lambdawtil_\ell - \Vwtil_Z.
\label{eq:Ttil_Z_cv}
\end{align}
Here elements of $\bfTwtil_Z$ can be written as
\begin{align}
\left[\,\bfTwtil_Z\,\right]_{\,ii}   &=
  \sum_{\ell \neq i}^L \alpha_\ell \Lambdawtil_\ell; \nonumber \\
 \left[\,\bfTwtil_Z\,\right]_{\,ij}  &=
      - [\,\bfMwtil\,]_{\,ij}  \qquad (i \neq j) \nonumber
      \\[1.3ex]
&= -\alpha_i^{1/2} \, \alpha_j^{1/2}\, \Lambdawtil_i^{1/2}
  \, \Lambdawtil_j^{1/2} \, c_{ij}
\label{eq:Ttil_elements}
\end{align}
where the coherence function $c_{ij}$ is defined to be
\begin{align}
 c_{ij} &\eqdef    \frac{\tr \left( \bfH_i^{\,\HT}\, \bfX_i \,
   \bfX_j^{\,\HT}\, \bfH_j \right) }
    { \tr \left( \bfH_{i\phantom{j}}^{\,\HT}\, \bfX_i \, \bfX_i^{\,\HT} \,
      \bfH_i  \right)^{1/2} \,
       \tr \left( \bfH_j^{\,\HT}\, \bfX_j \, \bfX_j^{\,\HT} \, \bfH_j
       \right)^{1/2}}  \nonumber \\
    \label{eq:coherence_def_orig}
\end{align}
Note that the coherence term is invariant to \emph{different} scalings
of the data $\bfX_\ell$ and $\bfX_n$, however $\bfTwtil_Z$ is not.
The matrix $\bfTwtil_Z$ is also a canonical form that is used
throughout this section
with the detectors and coefficients $\{ \alpha_\ell \}$ modified for the
particular noise model.

The maximization of $\Lambdawtil_Z$ with respect to the
channel gains is equivalent to minimizing the cross validation term in 
(\ref{eq:Ttil_Z_cv}) with respect to $\bfgwtil$.
It follows that, $\Vwtil_Z$ is equal to 
the smallest eigenvalue of $\bfTwtil_Z$. Then the GLR detector is
\begin{equation}
  \Lambdawtil_Z = \sum_{\ell=1}^L \alpha_\ell \Lambdawtil_\ell -
  \text{mineig} \left\{\,\bfTwtil_Z\,\right\}.
\end{equation}
The results of this section are summarized in panel $P_{21}$ in
Table~\ref{tbl:GLMdetectors}. 
Again, this conforms to the canonical forms of this paper.
Some insight can be obtained by considering some special cases.
\begin{center} \underline{\bf Example: Two Channels} \end{center}

\noindent In this section $\alpha_\ell = 1/L$ for all $\ell$, which
implies that the
overall detection statistic would just be scaled by $1/L$.
Therefore we ignore it in the following. For two channels we have
\begin{equation}
  \bfTwtil = \twomatstretch{ \Lambdawtil_2}
   {-\Lambdawtil_1^{1/2} \;\Lambdawtil_2^{1/2} \,c_{12} }
   {-\Lambdawtil_1^{1/2} \;
     \Lambdawtil_2^{1/2} \,c_{12}^{\,\mbox{\normalsize$\ast$}
   }}{\Lambdawtil_1}{1.5}.
\end{equation}
Denote the arithmetic and geometric means of the per-channel
detectors by  $A_{12} = (\Lambdawtil_1 + \Lambdawtil_2)/2$ and $G_{12}
= (\Lambdawtil_1 \, \Lambdawtil_2)^{1/2}$.  The cross-validation term
can be written as
\begin{equation}
  \Vwtil = \text{mineig}\{\,\bfTwtil\,\} = A_{12} - A_{12}\left( 1 +
  \frac{ G_{12}^2}{A^2_{12}} \left(\magsq{\,c_{12}\,}-1 \right) \right)^{1/2}.
\label{eq:two_channel_CVtil_def}
\end{equation}
The cross-validation term is
monotonically decreasing with respect to the coherence term
$\cct{12}$.  Again, this is intuitive.  As the coherence between each
channel's estimate of the mode amplitudes increases, the penalty
imposed by the cross-validation term decreases, which implies that the
overall detection statistic is larger.

There is additional insight one can obtain from this result. Let
\begin{equation}
  \Deltawtil_{12} = \frac{\Lambdawtil_1 - \Lambdawtil_2}{2}.
\end{equation}
Then, it can be shown that the square of the ratio of the geometric mean to the
arithmetic mean can be written as
\begin{equation}
  \left( \frac{\Gwtil_{12}}{\Awtil_{12}} \right)^2 =   1-\left(
  \frac{\Deltawtil_{12}}{\Awtil_{12}}
      \right)^2 \eqdef  1-\nu^{\,2}_{12}.
      \label{eq:coefficient_of_variation}
\end{equation}
Here $\nu_{12}^2 \in [0,1)$ is the squared
coefficient-of-variation of the detector statistics.  The larger
$\nu_{12}^2$, the larger the ``normalized-distance'' of the detector outputs.
One might expect that our confidence in the individual detector values
decreases when this coefficient increases, and this is so.
In this  two-channel case, the detector output is a function of the
arithmetic mean, the squared coefficient-of-variation of the
per-channel detector outputs, and the coherence of the matched filter
outputs on each channel.

\subsection{Unknown $\{ g_\ell \}$: common but unknown noise variance}

When the noise variances are unknown but identical, we can derive the
detector of this section by maximizing the CFAR detector in
(\ref{eq:Lambda_Z_CFAR}) with respect to the unknown channel-gains.
When the channels matrices are such that $\bfH_i^{\,\HT} \, \bfH_i =
\bfI_J$, we show in Appendix~\ref{sec:unknown_gains_appendix} that
the GLR detector can be written as
\begin{align}
  \Lambda_{Z,\text{CFAR}} &=
  \sum_{\ell=1}^L \alpha_\ell \Lambda_{\ell,\text{CFAR}} -
  \frac{ \bfg^{\,\HT}\, \bfT_{Z,\text{CFAR}} \, \bfg}{\bfg^{\,\HT}
    \, \bfg} \nonumber \\
  &\eqdef
  \sum_{\ell=1}^L \alpha_\ell \Lambda_{\ell,\text{CFAR}} -
  V_{Z,\text{CFAR}}.
\label{eq:Tcfar_Z_cv}
\end{align}
Under the conditions of this section,
\begin{equation}
  \alpha_\ell = \frac{\tr(\bfS_{\,\ell\ell})\EndStrut{0.55ex}}
  {\StartStrut{0.65ex}\tr(\SZZ)} \qquad \text{and} \qquad
  \Lambda_{\ell,\text{CFAR}} = 
  \frac{\tr(\bfP_{\,\bfH_\ell}\,\bfS_{\,\ell\ell})\EndStrut{0.55ex}}
  {\StartStrut{0.65ex}\tr(\bfS_{\,\ell\ell})}.
\label{eq:cfar_alpha}
\end{equation}
and the elements of $\bfT_{Z,\text{CFAR}}$ have the same structure as
those in (\ref{eq:Ttil_elements}) but the components are those
described in (\ref{eq:cfar_alpha}).  The procedures of
the previous section can be
duplicated to obtain the canonical detector
\begin{align}
  \Lambda_{Z,\text{CFAR}} &= \sum_{\ell=1}^L \alpha_\ell \,
  \Lambda_{\ell,\text{CFAR}} - \text{mineig}\{\,
  \bfT_{Z,\text{CFAR}} \, \} \nonumber \\
  &\eqdef
  \sum_{\ell=1}^L \alpha_\ell \,
  \Lambda_{\ell,\text{CFAR}} - V_{Z,\text{CFAR}}.
\end{align}
These results correspond to panel $P_{22}$ in
Table~\ref{tbl:GLMdetectors}.

\subsection{Unknown $\{ g_\ell \}$: different and unknown noise
  variances}

In this section, as was done in 
Section~\ref{sec:knownFZ_unknown_different_noise}, let
\begin{equation}
  \sigmawhat_\ell^2(0) = \frac{1}{N_\ell} \tr(\bfS_{\,\ell\ell}); \quad \text{and} \quad 
  \sigmawhat_\ell^2(1) = \frac{1}{N_\ell} \tr\left((\bfI-\bfP_{\bfH_\ell})\bfS_{\,\ell\ell} \right).
\end{equation}
We can then duplicate the procedures of the previous two sections to
obtain a detector structure
\begin{equation}
  \Lambda_{\{X_\ell\},\text{CFAR}}(\, \bfg\,) = \sum_{\ell=1}^L \alpha_\ell \,
  \Lambda^{(2)}_{\ell, \text{CFAR}} - 
  \frac{ \bfg^{\,\HT}\, \bfT_{\{X_\ell\},\text{CFAR}} \, \bfg}{\bfg^{\,\HT}
    \, \bfg} \nonumber \\
\end{equation}
where, in this case, $\alpha_\ell = N_\ell / N_Z$.  There is a
slight modification needed in the definition of the elements of the
matrix $\bfT_{\{ X_\ell\},\text{CFAR}}$ in this formula.  The off-diagonal elements
are
\begin{align}
  [ \bfT_{\{ \ell\},\text{CFAR}}]_{ij} & =
  \left(\frac{N_i \EndStrut{0.55ex}}{ \StartStrut{0.65ex} N_Z} \right)^{1/2} \, 
  \left(\frac{N_j \EndStrut{0.55ex}}{ \StartStrut{0.65ex} N_Z} \right)^{1/2} \,  \left( \frac{\tr(\bfP_{\bfH_i} \bfS_{ii}) \EndStrut{.55ex}}
  {\StartStrut{.65ex} \tr((\bfI-\bfP_{\bfH_i}) \bfS_{ii})}
  \right)^{1/2} 
 \left( \frac{\tr(\bfP_{\bfH_j} \bfS_{jj}) \EndStrut{.55ex}}
  {\StartStrut{.65ex} \tr((\bfI-\bfP_{\bfH_j}) \bfS_{jj})}
  \right)^{1/2}  c_{ij} \nonumber \\[1.5ex]
  & = \alpha_i^{1/2} \, 
  \alpha_j^{1/2} \, 
   F^{1/2}_{i,\text{CFAR}}\,
   F^{1/2}_{j,\text{CFAR}}\,c_{ij}.
\end{align}
Here $F_{i,\text{CFAR}}$ is an $F$-distributed random variable that
is used instead of the per-channel detectors
$\Lambda^{(2)}_{i,\text{CFAR}}$ to form the elements of $\bfT$. As
in the previous sections we optimize over $\bfg$ to obtain the GLR
detector
\begin{equation}
  \Lambda_{\{X_\ell\},\text{CFAR}} = \sum_{\ell=1}^L \alpha_\ell \,
  \Lambda^{(2)}_{\ell, \text{CFAR}} - 
  \text{mineig} \left\{\,\bfT_{\{X_\ell\},\text{CFAR}} \, \right\},
\end{equation}
which is a version of our canonical detector summarized in panel
$P_{23}$ in Table~\ref{tbl:GLMdetectors}.  As before, the subscript
``$\{ X_\ell \},\text{CFAR}$'' is used to indicate that the quantity is
invariant to different scalings of each channel's data.

\section{Unknown Channel-mode matrices $\bfH_\ell$}
\label{sec:unknown_channels}

In this section we consider cases where the channel matrices
$\{\,\bfH_\ell \,\}$ are unknown but constrained to have
rank $J$. In this case the channel gains,
whether they are known or unknown, can be ``absorbed'' into the
respective channel matrices.  Consequently the composite
channel matrix $\bfF$ in (\ref{eq:z_composite_model}) and
its whitened version $\bfFwtil = \RZZ^{-1/2} \, \bfF$ 
can be considered to be unknown. We assume that the number of measurement
vectors satisfies $M \geq J$.  

\subsection{Unknown $\bfF$: known noise variances}

When the noise variances are assumed known, we can use 
the clairvoyant detector in (\ref{eq:clairvoyant_form_comp})
\begin{equation}
  \Lambdawtil_Z =  \frac{1}{L}
  \tr \left( \bfP_{\, \bfFwtil }\,
  \SZZtil \right) 
  \label{eq:prelim_unknown_H_G}
\end{equation}
as a preliminary detector, with as yet unknown $\bfFwtil$.
This equation indicates that the detector is maximized when
the span of $\bfFwtil$ coincides with the span of the dominant subspace
of $\SZZtil$.  It follows that the
detector of (\ref{eq:prelim_unknown_H_G}), when compressed with this
estimate, consists of the sum of the dominant eigenvalues of the whitened,
composite, sample-covariance matrix. 
Denote the ordered eigenvalues of any $N \times N$
matrix $\bfK$ by $\lambda_1(\, \bfK \,) \geq
 \lambda_2(\, \bfK \,) \geq \cdots \geq \lambda_{N}(\, \bfK \,)$.
We can then write the detector of this section as
\begin{equation}
  \Lambdawtil_Z = \frac{1}{L}\sum_{n=1}^J \lambda_n(\, \SZZtil \, ).
  \label{eq:clairvoyant_unknown_A}
\end{equation}
Note that the covariance matrix in this expression 
consists of whitened data, {\it i.e.,} the detector consists of the
\mbox{variance-normalized} energy in the dominant subspace.
An equivalent form is
\begin{align}
  \Lambdawtil_Z &= \frac{1}{L} \left[ \tr \left(  \SZZtil  \right)
   - \sum_{n=J+1}^{N_Z} \lambda_n\left(\, \SZZtil \, \right) \right] \nonumber \\
   &\eqdef \frac{1}{L}\tr \left(  \SZZtil  \right)
  -\frac{1}{L}\SDEtil_Z
\end{align}
where $\widetilde{\calE}_Z$ represents the energy in the sub-dominant
subspace of $\SZZtil$.  Note that if $M < N_Z$, some of these
eigenvalues will be zero.  This expression can be expanded into our
canonical form
\begin{align}
 \Lambdawtil_Z
  &= \sum_{\ell=1}^{L} \frac{1}{L} \tr \left(  \bfSwtil_{\,\ell\ell}  \right) 
  -\frac{1}{L}\SDEtil_Z
  \nonumber \\
  &\eqdef \sum_{\ell=1}^{L} \alpha_\ell
  \left(\sum_{n=1}^J \lambda_n(\bfSwtil_{\,\ell\ell})  
  + \sum_{n=J+1}^{N_\ell} \lambda_n(\bfSwtil_{\,\ell\ell}) \right)
  -\frac{1}{L}\SDEtil_Z \nonumber \\
  &\eqdef \sum_{\ell=1}^{L} \alpha_\ell( \Lambdawtil_\ell +
  \SDEtil_\ell) -\frac{1}{L}\SDEtil_Z \nonumber \\
  &= \sum_{\ell=1}^{L} \alpha_\ell \Lambdawtil_\ell - 
    \left(\frac{1}{L}\SDEtil_Z - \sum_{\ell=1}^{L}\alpha_\ell \SDEtil_\ell \right)
  \nonumber \\
  &\eqdef \sum_{\ell=1}^{L} \alpha_\ell \Lambdawtil_\ell  - \Vwtil_Z,
  \label{eq:cv_form_unknown_H_G}
\end{align}
which adheres to the framework of the detectors in this paper.

The cross-validation term is the difference of the estimated power in
the per-channel ``noise'' subspaces  and the estimated power in the
noise subspace using the composite data.  Again, this is a measure of
how the per-channel results agree with the composite channel result.

Note that the eigenvalues of the sample covariance matrix are
\begin{align}
  \lambda_n\left(\SZZtil \right) 
  &= \frac{1}{M} \lambda_n \left(\bfZwtil \,  \bfZwtil^{\,\HT} \right) \nonumber \\
 &= \frac{1}{M} \lambda_n\left(\bfZwtil^{\,\HT} \, \bfZwtil \right) = \frac{1}{M}
  \lambda_n \left( \bfXwtil_1^{\,\HT} \, \bfXwtil_1\, +\,  \cdots\, +\, 
  \bfXwtil_L^{\,\HT} \, \bfXwtil_L \right). 
\end{align}
A Lidskii inequality \cite{Tao10}
\begin{equation}
  \sum_{n=1}^J \lambda_n(\bfA+\bfB) \leq \sum_{n=1}^J \lambda_n(\bfA)
  +\sum_{n=1}^J  \lambda_n(\bfB)    \quad (\bfA,\,\bfB \;
  \text{Hermitian})
\end{equation}
can be applied, recursively, to show
\begin{equation}
  \Lambdawtil_Z \leq \sum_{\ell=1}^L \Lambdawtil_\ell.
\end{equation}
It follows from this result that the cross-validation term is
non-negative.  The detector results are summarized in panel $P_{31}$
of Table~\ref{tbl:GLMdetectors}.
Some insight can be obtained by considering the following simple case.

\medskip
\begin{center} \underline{\bf Rank-one signal, M=2 observations} \end{center}
Suppose that the channel matrix $\bfF$ is rank-one and that two
observations have been made: $\bfZ = [\, \bfz[1] \; \bfz[2] \,] \in
\C^{N_Z \times 2}$.  From (\ref{eq:clairvoyant_unknown_A}) we have the
multi-channel composite detector
\begin{align}
  \Lambdawtil_Z &= \frac{1}{L}\lambda_1(\SZZtil) = \frac{1}{L}\text{maxev}
  \twomatstretch{\bfzwtil[1]^\HT \, \bfzwtil[1]}
      {\bfzwtil[1]^\HT \, \bfzwtil[2]}
      {\bfzwtil[2]^\HT \, \bfzwtil[1]}
      {\bfzwtil[2]^\HT \, \bfzwtil[2]}{1.4} \nonumber \\
      &= \frac{1}{L}\left( \frac{ \bfzwtil[1]^\HT \, \bfzwtil[1]
        + \bfzwtil[2]^\HT \, \bfzwtil[2]}{2} +
      \frac{1}{2}D^{1/2}\left(\SZZtil \right) \right)
\label{eq:rank1_lambda_Z}
\end{align}
where the discriminant is 
\begin{align}
  &D(\SZZtil) = 
      \left(\,  \bfzwtil[1]^\HT \, \bfzwtil[1]
      + \bfzwtil[2]^\HT \, \bfzwtil[2]\, \right)^2   + 4 
      \left(\,  \bfzwtil[1]^\HT \, \bfzwtil[1]\, \right)
        \left(\,  \bfzwtil[2]^\HT \, \bfzwtil[2]\,
        \right) \left(\magsq{\,c(\SZZtil)\,} - 1 \right)
\end{align}
and 
\begin{equation}
        \magsq{\,c(\SZZtil)\,} \eqdef \frac{\big\vert \, 
         \bfzwtil[1]^\HT \, \bfzwtil[2]\,\big\vert^{\, 2}}
      {\left( \, \bfzwtil[1]^\HT \, \bfzwtil[1]\, \right)^{1/2} \, 
      \left( \, \bfzwtil[2]^\HT \, \bfzwtil[2]\, \right)^{1/2} }
\end{equation}
is the coherence between the data measurements.  We can write
(\ref{eq:rank1_lambda_Z}) as
\begin{align}
  \Lambdawtil_Z &= \frac{1}{L} \sum_{\ell=1}^L  \frac{ 
      \bfxwtil_\ell[1]^\HT \, \bfxwtil_\ell[1]
       + \bfxwtil_\ell[2]^\HT \, \bfxwtil_\ell[2] }{2}
       + \frac{1}{2L}D^{1/2}(\SZZtil) \nonumber \\
       &= \sum_{\ell=1}^L \frac{1}{L} \lambda_1(\bfSwtil_{\,\ell\ell}) - 
         \frac{1}{2L}\left[ \sum_{\ell=1}^L D^{1/2}(\bfSwtil_{\,\ell\ell}) -
         D^{1/2}(\SZZtil) \right] \nonumber \\
       &\eqdef \sum_{\ell=1}^L \alpha_\ell \Lambdawtil_\ell -
         \Vwtil_Z.
\end{align}
This has our canonical structure.  Suppose that the data vectors
$\bfz[m]$ are colinear, {\it i.e.,} their squared coherence is one.
It follows that under these conditions
\begin{align}
  D^{1/2}(\SZZ) &= 
      \bfzwtil[1]^\HT \, \bfzwtil[1]\, + 
      \bfzwtil[2]^\HT \, \bfzwtil[2] \nonumber \\
      &= \sum_{\ell=1}^L 
      \bfxwtil_\ell[1]^\HT \, \bfxwtil_\ell[1]\, + 
      \bfxwtil_\ell[2]^\HT \, \bfxwtil_\ell[2]
\end{align}
and
\begin{equation}
  D^{1/2}(\bfSwtil_{\,\ell\ell}) = 
      \bfxwtil_\ell[1]^\HT \, \bfxwtil_\ell[1]\, + 
      \bfxwtil_\ell[2]^\HT \, \bfxwtil_\ell[2].
\end{equation}
It follows that the cross-validation term is zero.  Now suppose that,
for each channel, the data vectors are orthogonal:
$\big\vert \, c(\bfSwtil_{\,\ell\ell})\, \big\vert^{\,2} = 0$.
Then $\big\vert \, c(\SZZ)\, \big\vert^{\,2} = 0$ as well.  Now the 
cross-validation term  
\begin{align}
   &\Vwtil_Z = \frac{1}{2} \left( \sum_{\ell=1}^L \frac{1}{L} \bigg\vert
   \, \bfxwtil_\ell[1]^\HT \, \bfxwtil_\ell[1]\, - 
      \bfxwtil_\ell[2]^\HT \, \bfxwtil_\ell[2] \, \bigg\vert \right. \left. -
      \frac{1}{L} \bigg\vert \,  \bfzwtil_\ell[1]^\HT \, \bfzwtil_\ell[1]\, - 
      \bfzwtil_\ell[2]^\HT \, \bfzwtil_\ell[2] \, \bigg\vert \right)
\end{align}
is a function of the variation in the energy in each snapshot.
Between these two extremes the cross-validation term is a function of
both the coherence of the data and the variation in the energy of the
data vectors. This
corresponds to the results in Section~\ref{sec:Unknown_channel_gains}.

\subsection{Unknown $\bfF$: common but unknown noise variance}

In the case where the noise variances are unknown, but assumed to be
the same, we can begin the development of the GLR detector of this
section with the CFAR detector in
(\ref{eq:Lambda_Z_CFAR})
\begin{equation}
  \Lambda_{Z,\text{CFAR}} =
  \frac{\tr \left(\bfP_{\,\bfF}\, \SZZ\right)}{\tr
    \left(\SZZ\right)}.
\end{equation}
The ML estimate of the composite
subspace is the dominant subspace of composite covariance matrix
$\SZZ$, where now the data covariance
consists of \emph{non-whitened} data.  Using the results of the
previous section and those in Section~\ref{sec:idealized_CFAR},
it follows that the CFAR detector in this case is
\begin{align}
  \Lambda_{Z,\text{CFAR}} &=
  \frac{ \sum_{n=1}^J
    \lambda_n(\, \SZZ\, ) \EndStrut{0.9ex}}
  {   \StartStrut{0.9ex} \tr(\SZZ) }.
  \label{eq:CFARZ_unknown_F}
\end{align}
This detector is invariant to a scaling of the composite data $\bfZ$.
Following the procedures of the previous section and those in
Section~\ref{sec:idealized_CFAR} we can write the
detector in (\ref{eq:CFARZ_unknown_F}) as
\begin{align}
  &\Lambda_{Z,\text{CFAR}} =
  \frac{1}{\tr \left(\SZZ\right)} \big\{ \tr \left(\SZZ\right)  - 
   \sum_{n=J+1}^{N_Z} \lambda_n(\, \SZZ\, ) \, 
   \big\} \nonumber \\
   &\quad = 1 - \SDE_Z/ \left( \tr\left( \SZZ \right) \right)
   \eqdef 1 - \SDE_{Z,\text{CFAR}} \nonumber \\
   &\quad = \sum_{\ell=1}^{L}
   \frac{\tr \left( \bfS_{\,\ell\ell} \right)}{\tr \left( \SZZ \right)}
             - \SDE_{Z,\text{CFAR}}  
   = \sum_{\ell=1}^{L} \alpha_\ell - \SDE_{Z,\text{CFAR}}\nonumber \\
   &\quad= \sum_{\ell=1}^{L} \alpha_\ell\,(1-\SDE_{X_\ell,\text{CFAR}})
  -\left(\SDE_{Z,\text{CFAR}} - \sum_{\ell=1}^{L}
             \alpha_\ell\SDE_{X_\ell,\text{CFAR}}\right)  \nonumber \\
   &\eqdef \sum_{\ell=1}^{L} \alpha_\ell \Lambda_{X_\ell,\text{CFAR}} - \CVZcfar
\end{align}
The cross-validation term now uses data-normalized energy in the
sub-dominant subspaces of each channel and of the composite channel.
It is invariant to a scaling of the composite data $\bfZ$. See panel
$P_{23}$ in Table~\ref{tbl:GLMdetectors} for a summary.
  
\subsection{Unknown $\bfF$: different and unknown noise variances}

In this section, as in Section~\ref{sec:knownFZ_unknown_different_noise},
we use the approximate per-channel noise estimates under $H_1$ 
\begin{equation}
 \sigmawhat_\ell^2(1) = \frac{1\EndStrut{0.55ex}}{\StartStrut{0.65ex}
   N_\ell} \tr( \, (\bfI - \bfP_{\,\bfH_\ell}) \, \bfS_{\,\ell\ell}).
\end{equation}
It follows that, prior to estimating the channel matrices, we can
start with the detector of that section
\begin{align}
  &\Lambdawtil_Z(\{\bfH_\ell\},\bfF) = 
       \sum_{\ell=1}^{L} 
  \frac{N_\ell\EndStrut{0.55ex}}{\StartStrut{0.65ex} N_Z }
  \ln \left( 
  \frac{\tr(\bfS_{\,\ell\ell})\EndStrut{0.55ex}}{\StartStrut{0.65ex}
    \tr( \, (\bfI - \bfP_{\bfH_\ell})\,\bfS_{\,\ell\ell})} \right)
 - 
  \left( \sum_{\ell=1}^L\frac{N_\ell\EndStrut{0.55ex}}{\StartStrut{0.65ex} N_Z }  
  \;\frac{\tr(\bfP_{\bfH_\ell}\,\bfS_{\,\ell\ell})\EndStrut{0.55ex}}{\StartStrut{0.65ex}
    \tr( \, (\bfI - \bfP_{\bfH_\ell})\,\bfS_{\,\ell\ell})} \right.  \left. -
  \frac{1\EndStrut{0.55ex}}{\StartStrut{0.65ex} N_Z}
  \tr(\bfP_{\bfF} \, \bfSwtil_{ZZ}(\{\,\sigmawhat_\ell^2(1)\,\} )\right)
\label{eq:detector_pre_Hest}
\end{align}
where we have expanded the cross-validation term.  Even though
$\bfF$ is a function of $\{\bfH_\ell\}$ we will estimate it separately
from $\{\bfH_\ell\}$ 
so that the per-channel detectors use only their local data.
Let the eigenvalue decomposition
of the per-channel sample covariance matrix be denoted by
\begin{equation}
  \bfS_{\,\ell\ell} = \left[ \, \bfU_\ell \:\: \bfV_\ell \, \right] \,
  \bfLambda_\ell \,
  \twovecstretch{\bfU_\ell^{\,\HT}}{\bfV_\ell^{\,\HT}}{1.3}
\end{equation}
where $\bfU_\ell \in \C^{N_\ell \times J}$ are the dominant
eigenvectors of $\bfS_\ell$.
It is not difficult to show that the per-channel ML estimate of the span
of $\bfH_\ell$ is
$ \subspace{\, \bfHwhat_{\ell} \, } = \subspace{ \, \bfU_\ell \,}$.
Compressing the detector in (\ref{eq:detector_pre_Hest}) with this
estimate yields
\begin{align}
  &\Lambdawtil_Z(\bfF) = 
       \sum_{\ell=1}^{L} 
  \frac{N_\ell\EndStrut{0.55ex}}{\StartStrut{0.65ex} N_Z}
  \ln \left( 1+
  \frac{\sum_{n=1}^{N_\ell} \lambda_n(\bfS_{\,\ell\ell})\EndStrut{0.55ex}}
       {\StartStrut{0.65ex}
      \sum_{n=J+1}^{N_\ell}\lambda_n(\bfS_{\,\ell\ell})} \right)
 - 
  \left( \sum_{\ell=1}^L 
  \frac{N_\ell \EndStrut{0.55ex}}{\StartStrut{0.65ex} N_Z}
  \frac{\sum_{n=1}^J \lambda_n(\bfS_{\,\ell\ell})\EndStrut{0.55ex}}
       {\StartStrut{0.65ex}
     \sum_{n=J+1}^{N_\ell} \lambda_n(\bfS_{\,\ell\ell})}
       \right. \left.
  -\frac{1\EndStrut{0.55ex}}{\StartStrut{0.65ex}N_Z}
     \tr(\bfP_{\bfF} \, \bfSwtil_{ZZ}(\{\,\sigmawhat_\ell^2(1)\,\}
    ))\right) \nonumber \\
     &\eqdef \sum_{\ell=1}^{L} \alpha_\ell\, \Lambda_\ell - V_{
       \{X_\ell\},\text{CFAR}}(\,\bfF\,).
\end{align}
It remains to estimate $\subspace{\bfF}$.  It is again easy to show
that $\subspace{ \,\bfFwhat\, }$ is the span of the $J$ dominant
eigenvectors of $\bfSwtil(\, \{ \, \sigmawhat_\ell^2(1) \, \}\,)$.
Then the cross-validation term is
\begin{align}
\bfV_{\{ X_\ell \},\text{CFAR}} 
&= \sum_{\ell=1}^L
  \frac{N_\ell \EndStrut{0.55ex}}{\StartStrut{0.65ex} N_Z }
  \frac{\sum_{n=1}^J \lambda_n(\bfS_{\,\ell\ell})\EndStrut{0.55ex}}
       {\StartStrut{0.65ex}
     \sum_{n=J+1}^{N_\ell} \lambda_n(\bfS_{\,\ell\ell})}
    - \sum_{j=1}^J
    \lambda_j \ \bfSwtil_{ZZ}\left( \left\{\,\sigmawhat_\ell^2(1)\, \right\} \right) 
    \nonumber \\
    &=\sum_{\ell=1}^L \alpha_\ell \phi_{\ell,\text{CFAR}} -
    \calE_{\{X_\ell\},\text{CFAR}}
\end{align}


\begin{table*}[p]
  \begingroup
  \renewcommand{\arraystretch}{1.4}
    \newcommand{\llen}{.9ex}
  \setlength{\tabcolsep}{\llen}

    \newcommand{\ColOneStartPad}{\rule{2.0ex}{0pt}}
    \newcommand{\ColOneEndPad}{\rule{0.0ex}{0pt}}

    \newcommand{\ColTwoStartPad}{\rule{2.0ex}{-10pt}}
    \newcommand{\ColTwoEndPad}{\rule{0.0ex}{0pt}}

    \newcommand{\ColThreeStartPad}{\rule{2.5ex}{0pt}}
    \newcommand{\ColThreeEndPad}{\rule{0.0ex}{0pt}}

    \newcommand{\ColFourStartPad}{\rule{1.0ex}{0ex}}
    \newcommand{\ColFourEndPad}{\rule{1.0ex}{0pt}}

  \begin{center}

    \begin{tabular}{|c|c|c|c|c|c|c|}

      \multicolumn{7}{c}{\large {Detector Structure:
        $\Lambda_Z = \sum_{\ell=1}^{L}\alpha_\ell\, \Lambda_\ell  - \cvsymb$}}
        \\[4ex] \cline{5-7}

       \multicolumn{4}{c|}{}
       & \multicolumn{3}{c|}{ Noise Models}
        \\ \cline{5-7}
        
       \multicolumn{4}{c|}{}
       & {\StartStrut{1.5ex}$\{\sigma_\ell^2\}$ known}
       & {$\{\sigma_\ell^2\} = \sigma^2_Z$ unknown}
       & {$\{\sigma_\ell^2\}$ unknown}
        \\ \cline{1-7}

       \multicolumn{4}{|c|}{scaling term}
       & {$\StartStrut{1.7ex} \alpha_\ell =
           \frac{1\EndStrut{0.5ex}}{\StartStrut{0.7ex}L}$}
       & {$\alpha_\ell = \frac{\tr(\bfS_{\,\ell\ell})\EndStrut{0.5ex}}
              {\StartStrut{0.7ex} \tr(\SZZ)}$}
       & {$\alpha_\ell = \frac{N_\ell\EndStrut{0.5ex}}{\StartStrut{0.7ex}
             N_Z}\EndStrut{2ex}$}
        \\ \cline{1-7}

       \multicolumn{4}{|c|}{$\Lambda_Z$ scaling invariances}
       &  none
       &  $\bfZ$
       &  $\bfX_\ell$, independently
          \\ \cline{1-7}

          \multicolumn{4}{|c|}{
            \begin{tabular}{c}
              per-channel detector:\\
              known $\bfH_\ell$ 
            \end{tabular} }
       & {$\StartStrut{1.7ex} \Lambda_\ell =
         \frac{\tr(\,\bfP_{\bfH_\ell}\,\bfS_{\,\ell\ell}\,)
           \EndStrut{0.6ex}}{\StartStrut{0.8ex}\sigma_\ell^2}
         $}
       & {$\Lambda_\ell =
         \frac{\tr(\,\bfP_{\bfH_\ell}\,\bfS_{\,\ell\ell}\,)
           \EndStrut{0.6ex}}{\StartStrut{0.7ex} \tr(\,\bfS_{\,\ell\ell}\,)}
         $}
       & {$\Lambda_\ell = \ln \left( 1+
           \frac{\tr(\,\bfP_{\bfH_\ell}\,\bfS_{\,\ell\ell}\,)\EndStrut{0.6ex}}
                {\StartStrut{0.7ex}
               \tr(\, (\bfI - \bfP_{\bfH_\ell}) \, \bfS_{\,\ell\ell}\, )}\right)
             \EndStrut{2ex}$}
       \\  \cline{1-7}


        \multicolumn{4}{|c|}{\begin{tabular}{c}per-channel detector:\\
            unknown, rank-$J$ $\bfH_\ell$ \end{tabular} }
        & {$\StartStrut{1.7ex}
   \Lambda_\ell = \frac{\sum_{j=1}^{J} \lambda_j(\,
     \bfS_{\,\ell\ell})\EndStrut{0.8ex}}{\StartStrut{0.8ex}\sigma_\ell^2} $}
        & {$\Lambda_\ell =
            \frac{\sum_{j=1}^{J} \lambda_j(\, \bfS_{\,\ell\ell}\,
              )\EndStrut{0.8ex}}
                 {\StartStrut{0.8ex}\tr(\,\bfS_{\,\ell\ell}\,) } $}
        & {$\Lambda_\ell = \ln \left( 1 + 
      \frac{\sum_{j=1}^{J} \lambda_j(\, \bfS_{\,\ell\ell}\,
        )\EndStrut{0.8ex}}{\StartStrut{0.8ex}\sum_{j=J+1}^{N_\ell}
        \lambda_j(\, \bfS_{\,\ell\ell} \, ) } \right)
             \EndStrut{2ex}$}
       \\ \cline{1-7}

       \multicolumn{7}{c}{\StartStrut{1ex}} \\ \cline{1-7}


    
       & \ColTwoStartPad
           \begin{rotate}{90} \hspace*{-5.5ex}
             $g_\ell$ known
           \end{rotate}
         \ColTwoEndPad
       & \ColThreeStartPad
         \begin{rotate}{90} \hspace*{-8.5ex}
           \begingroup
           \renewcommand{\arraystretch}{0.8}
           \begin{tabular}{c}
             $\tr(\bfH_\ell^{\,\HT}\,\bfH_\ell) = J$ \\
           \end{tabular}
           \endgroup
         \end{rotate}
         \ColThreeEndPad
       & \ColFourStartPad
         \begin{rotate}{90} \hspace*{-9ex}
           metric: Euclidean
         \end{rotate}
         \ColFourEndPad
         & \ReallyBigTstrut \begin{tabular}{c}
           $V\eqdef\Vwtil_Z =
           \frac{1\EndStrut{0.55ex}}{\StartStrut{0.65ex}L}
           \sum_{p=1}^P \Vwtil_{Z_p}$ \\
             $\Vwtil_{Z_p} = \tr( \bfQwtil^{-1}_{E_pE_p} \,
           \bfSwtil_{E_pE_p})$ \\[2ex]
           Eqs. (\ref{eq:PhiZZ_canonical}) and
           (\ref{eq:recursive_partition_cv})
           \end{tabular}
       & {$V\eqdef V_{Z,\text{CFAR}} =
        \frac{L\,\EndStrut{0.55ex}}{\StartStrut{0.65ex} \tr( \SZZ)}
        \Vwtil_Z \, \bigg\vert_{\sigma_\ell^{\,2} = 1}$}
       & \begin{tabular}{c}  {$V \eqdef V_{\{X_\ell\},\text{CFAR}} =
           \frac{L}{\StartStrut{0.65ex}N_Z} \Vwtil_Z$}  \\[1.3ex]
          {{evaluated at }
            $\sigma_\ell^{\,2}=
            \frac{\tr((\bfI-\bfP_{\bfH_\ell}\EndStrut{.55ex})
              \bfS_{\,\ell\ell})}{\StartStrut{.65ex} N_\ell}$}
         \end{tabular}
           \\[10ex] \cline{2-7}

    
           \ColOneStartPad
           \begin{rotate}{90} 
             \hspace*{4ex} Known $\bfH_\ell$
           \end{rotate}
           \ColOneEndPad
         & \ColTwoStartPad
           \begin{rotate}{90} 
             \hspace*{-5.8ex} $g_\ell$ unknown
           \end{rotate}
           \ColTwoEndPad
         & \ColThreeStartPad
           \begin{rotate}{90}
             \hspace*{-6ex} $\bfH_\ell^\HT \, \bfH_\ell = \bfI_J$
           \end{rotate}
           \ColThreeEndPad
         & \ColFourStartPad
           \begin{rotate}{90}
             \hspace*{-8ex} metric: coherence
           \end{rotate}
           \ColFourEndPad
           & \ReallyBigTstrut
           \begin{tabular}{c}
           {$\cvsymb = \text{mineig}\{ \, \bfTwtil_Z
           \,\}$} \\[2.ex] Eqs (\ref{eq:Ttil_elements})
     and (\ref{eq:coherence_def_orig})
           \end{tabular}
           & \begin{tabular}{c}
               $\cvsymb =
           \text{mineig}\{\,\bfT_{Z,\text{CFAR}}\,\}$ \\[2ex]
                {$\bfT_{Z,\text{CFAR}} =
            \frac{L\EndStrut{.55ex}}{\StartStrut{.65ex}\tr(\SZZ)} \bfTwtil_Z
              \bigg\vert_{\sigma_\ell^{\,2} = 1}$}
            \end{tabular}
         & \begin{tabular}{c} $\cvsymb =
               \text{mineig}\{\,\bfT_{\{ \bfX_\ell \},
                 \text{CFAR}}\,\}$ \\[2ex]
            {$\bfT_{\{X_\ell\},\text{CFAR}} =
              \frac{L\EndStrut{.55ex}}{\StartStrut{.65ex}N_Z}
              \bfTwtil_Z$}\\[2ex]
            {{evaluated at }
            $\sigma_\ell^{\,2}=
            \frac{\tr((\bfI-\bfP_{\bfH_\ell}\EndStrut{.55ex})
              \bfS_{\,\ell\ell})}{\StartStrut{.65ex} N_\ell}$}
               \end{tabular}
           \\[10ex] \cline{1-7}

    
             \ColOneStartPad
             \begin{rotate}{90}
               \hspace*{-7.5ex} Unknown $\bfH_\ell$
             \end{rotate}
             \ColOneEndPad
           & \ColTwoStartPad
             \begin{rotate}{90}
               \hspace*{-7.5ex} $g_\ell$ (un) known
             \end{rotate}
             \ColTwoEndPad
           & \ColThreeStartPad
             \begin{rotate}{90}
               \hspace*{-6.0ex} $\bfH_\ell$ rank-$J$
             \end{rotate}
             \ColThreeEndPad
           & \ColFourStartPad
             \begin{rotate}{90}
                    \hspace*{-10.8 ex} metric: subspace energy
             \end{rotate}
             \ColFourEndPad
           & \ReallyBigTstrut 
             $\cvsymb =  \SDE_Z -
             \sum_{\ell=1}^{L}\alpha_\ell\SDE_\ell $
              & \begin{tabular}{c}$\cvsymb = $\\
                $ \SDE_{Z,\text{CFAR}}-\sum_{\ell=1}^{L}\alpha_\ell\SDE_{X_\ell,\text{CFAR}}$
             \end{tabular}
           & \begin{tabular}{c} 
                $V=V_{\{ X_\ell \},\text{CFAR}}$  \\
                $=\sum_{\ell=1}^L \alpha_\ell \phi_{\ell,\text{CFAR}} -
    \calE_{\{X_\ell\},\text{CFAR}} $
                 \end{tabular}
           \\[10ex] \cline{1-7}

    \end{tabular}
  \end{center}
  \endgroup
  \caption{Detectors For General Linear Model}
  \label{tbl:GLMdetectors}
\end{table*}

\section{Conclusion}

In this paper we have developed a general framework for describing GLR detectors
for multi-channel problems.  The results are general and many existing
multi-channel detectors, as well as the new detectors developed here, can be
written in the canonical form of this paper. The structure of the  composite
detector consists of a weighted combination of the
detectors for each individual channel modified by a cross-validation
or fusion term.  The cross-validation term is a measure of the
concordance between the detectors for the individual channels.  If
the agreement between the per-channel detectors is low, then the
composite detector
statistic is reduced to indicate the reduced confidence in the
per-channel results.  The cross-validation term is a function of the models for both
 noise and channel, and often has a intuitive
interpretation.  For example, it can be a function of the 
difference in the local  per-channel estimates of the signal
amplitudes.  In other instances, the cross-validation term is a
function of the coherence of these amplitude estimates and extent of
the dispersion of the per-channel detector outputs.

The framework and results are applicable, for example, to any problem
that uses interferometry to infer the presence of a source
(e.g. seismology, cosmology, multi-static radar/sonar).   
Another potential application is to the detection of a band-limited signal (e.g a communication signal) 
embedded in multiple time series.
In future work we intend to further develop the approach for source localization problems.

\appendix
\section{Appendix}
\subsection{Decomposition of composite detectors: form 1 }
\label{sec:MultiChan_decomp}
  
The results of this section are based on 
alternate forms of the matrix quadratic form
\begin{equation}
  \bfPhi_{ZZ} = \bfZ^{\,\HT} \, \bfP_{\bfF} \, \bfZ
\label{eq:PhiZZ_def}
\end{equation}
where at this point we only partition the model into two channels 
\begin{equation}
  \bfF = \twovec{\bfF_X}{\bfF_Y}; \quad \bfZ = \twovec{\bfX}{\bfY},
\end{equation}
each of which can be composite.
Denote the ML estimate of the signal amplitudes using \emph{only} the
data from channel-$X$ by
\begin{align}
  \AXhat &= (\bfF_X^{\,\HT} \, \bfF_X)^{-1} \,
  \bfF_X^{\,\HT} \, \bfX 
 \quad  \sim \calC\calN(\bfA,(\bfF_X^{\,\HT} \, \bfF_X )^{-1}) 
  \eqdef \calC\calN(\bfA,\QXX).
\end{align}
Using this definition allows us to write the inverse contained in the
projection matrix as
\begin{align}
  \left(\bfF^{\,\HT}\,\bfF \right)^{-1} &=
  \left(  \bfF_X^{\,\HT} \, \bfF_X\,
  +  \, \bfF_Y^{\,\HT} \, \bfF_Y \right)^{-1} \nonumber
  \\
  &= \left( \QXX^{-1} + \QYY^{-1} \right)^{-1} \eqdef \QZZ
\end{align}
where $\QZZ$ is the error covariance of $\AZhat$.

The matched-filter portion of the quadratic form is
\begin{equation}
  \bfF^{\,\HT} \; \bfZ =   \bfF_X^{\,\HT} \,
  \bfX\;+ \; \bfF_Y^{\,\HT} \, \bfY.
\end{equation}
The first term in this expression can be expanded into
\begin{align}
  \bfF_X^{\,\HT} \, \bfX
  &= \left( \bfF_X^{\,\HT} \, \bfF_X \right) \, \left[
  \left( \bfF_X^{\,\HT} \, \bfF_X
  \right)^{-1} \,\bfF_X^{\,\HT} \, \bfX \right]\nonumber \\
  &= \QXX^{-1} \, \AXhat.
  \label{eq:Xtil_matched_filter}
\end{align}
At this point the quadratic form $\bfPhi_{ZZ}$ is equivalent to
\begin{align}
  \bfPhi_{ZZ} &=  \bfZ^{\,\HT} \,
  \bfP_{\bfF} \, \bfZ \nonumber \\
  &= 
  \left( \QXX^{-1} \, \AXhat + \QYY^{-1} \, \AYhat \right)^{\,\HT}
  \left( \QXX^{-1} + \QYY^{-1} \right)^{-1} \;  \left( \QXX^{-1} \, \AXhat + \QYY^{-1} \, \AYhat \right)
  \nonumber \\
  &\eqdef \bfPhi_{XX} + \bfPhi_{XY} + \bfPhi_{YX} + \bfPhi_{YY}.
\label{eq:PhiZZ_expansion}
\end{align}
A cross term can be expressed as
\begin{align}
  \bfPhi_{XY} &= \AXhat^{\,\HT}\,\QXX^{-1} 
  \left( \QXX^{-1} + \QYY^{-1} \right)^{-1} \QYY^{-1} \, \AYhat \nonumber
  \\
  &= \AXhat^{\,\HT}\, 
  \left( \QXX + \QYY \right)^{-1}  \AYhat \nonumber \\
  &= \AXhat^{\,\HT}\, \QEEinv \, \AYhat
\label{eq:PhiXY_def}
\end{align}
where it can be shown that
\begin{equation}
  \QEE = \QXX + \QYY
  \label{eq:QEE_def}
\end{equation}
is the covariance matrix of $\AXhat - \AYhat$.
The matrix inversion lemma is now used to write a quadratic term in
(\ref{eq:PhiZZ_expansion}) as
\begin{align}
  \bfPhi_{XX} &= \AXhat^{\,\HT} \, \QXX^{-1} \, 
  \left( \QXX^{-1} + \QYY^{-1} \right)^{-1} \QXX^{-1} \, \AXhat
  \nonumber \\
  &= \AXhat^{\,\HT} \, \QXX^{-1} \, \left(\QXX -
  \QXX(\QXX + \QYY)^{-1}
  \QXX\right) \; 
 \QXX^{-1} \AXhat \nonumber \\[2ex]
  &=\AXhat^{\,\HT} \, \QXX^{-1} \, \AXhat -
  \AXhat^{\,\HT} \, (\QXX+\QYY)^{-1} \, \AXhat \nonumber \\[2ex]
  &= \bfX^{\,\HT} \, \bfP_{\bfF_X} \, \bfX -
   \AXhat^{\,\HT} \, \QEEinv \, \AXhat.
\label{eq:PhiXX_def}
\end{align}
It follows from (\ref{eq:PhiXY_def}) and (\ref{eq:PhiXX_def}) that
$\bfPhi_{ZZ}$ has an alternate form
\begin{align}
  \bfZ^{\,\HT} \, \bfP_{\bfF} \, \bfZ
 &  = \bfX^{\,\HT} \, \bfP_{\bfF_X} \, \bfX \;+\;
  \bfY^{\,\HT} \, \bfP_{\bfF_Y} \, \bfY  - (\AXhat -
   \AYhat)^{\,\HT}\, \QEEinv \, (\AXhat -\AYhat) \nonumber \\[1 ex]
   &=
  \bfX^{\,\HT} \, \bfP_{\bfF_X} \, \bfX \;+\;
  \bfY^{\,\HT} \, \bfP_{\bfF_Y} \, \bfY\; -\;
  M\,\tr(\bfQ^{-1}_{EE} \bfE\bfE^{\,\HT}/M) \nonumber \\
   &=
  \bfX^{\,\HT} \, \bfP_{\bfF_X} \, \bfX \;+\;
  \bfY^{\,\HT} \, \bfP_{\bfF_Y} \, \bfY\; -\;
  M\,\tr( \bfQ^{-1}_{EE} \bfS_{\,EE}).
   \label{eq:PhiZZ_canonical}
\end{align}
Suppose that we recursively partition each composite channel until the
final partitions each consist of a single channel (i.e. they are not
composite). The identity in (\ref{eq:PhiZZ_canonical}) can be applied
to each partitioning step in this process.
Denote the total number of partitioning steps within this process by
$P$. Let $p$ be an index into a table of $P$ entries, with each
entry describing the parameters of the corresponding partitioning.  In
particular let $Z_p$ denote the total number of channels in a
partition before it is divided. Then (\ref{eq:PhiZZ_canonical})
can be expanded into
\begin{align}
  \bfZ^{\,\HT} \, \bfP_{\bfF} \, \bfZ &=
  \sum_{\ell=1}^L \bfX_\ell^{\,\HT} \, \bfP_{\bfF_\ell} \, \bfX_\ell
  - M\sum_{p=1}^P \tr ( \bfQ^{-1}_{E_p E_p}\bfS_{E_pE_p}) \nonumber \\
  &\eqdef \sum_{\ell=1}^L \bfX_\ell^{\,\HT} \, \bfP_{\bfH_\ell} \, \bfX_\ell
  - M\bfV
\label{eq:recursive_partition_cv}
\end{align}
where we have used the fact that $\bfP_{\bfF_\ell} = \bfP_{g_\ell
  \bfH_\ell} = \bfP_{\bfH_\ell}$.
This identity can be used to write the detector
\begin{align}
  \Lambda_{Z,\text{CFAR}} &=
  \frac{1}{\tr(\SZZ)} \tr \left( \bfP_{\bfF} \, \SZZ \right) \nonumber
  \\ &=
  \frac{1}{M\tr(\SZZ)} \tr \left( \bfZ^{\,\HT} \,
  \bfP_{\bfF} \, \bfZ \right) \nonumber \\
   &= \frac{1}{M\tr(\SZZ)}\sum_{\ell=1}^L\tr
  \left(\bfX_\ell^{\,\HT} \, \bfP_{\bfH_\ell} \, \bfX_\ell\right)
-\frac{1}{\tr(\SZZ)} \sum_{p=1}^P \tr\left( \bfQ^{-1}_{E_pE_p}
   \bfS_{E_pE_p}\right)
   \nonumber \\
   &= \sum_{\ell=1}^L\frac{\tr(\bfS_{\,\ell\ell})}{\tr(\SZZ)} 
    \frac{\tr \left( \bfP_{\bfH_\ell} \, \bfS_{\,\ell\ell}
      \right)}{\tr(\bfS_{\,\ell\ell})} - V_{Z,\text{CFAR}} \nonumber \\
   &\eqdef \sum_{\ell=1}^L \alpha_\ell \, \Lambda_{X_\ell,\text{CFAR}} - V_{Z,\text{CFAR}}
\label{eq:cross_validation_term_1}
\end{align}
This is an instance of the canonical detector structure described in
this document.

\subsection{Decomposition of multi-channel detectors: form 2 }
\label{sec:MultiChan_decomp_form2}

A second form for the cross validation matrix $V_{Z,\text{CFAR}}$  in
(\ref{eq:cross_validation_term_1}) can be found by
expanding the detector in this expression as
\begin{align}
  \Lambda_{Z,\text{CFAR}} &=
  \frac{1\EndStrut{0.55ex}}{\StartStrut{0.65ex} M \tr(\SZZ)}
    \tr \left( \bfZ^{\,\HT} \, \bfP_{\bfF} \, \bfZ  \right) 
  = \frac{1\EndStrut{0.55ex}}{\StartStrut{0.65ex} M\tr(\SZZ)}
  \sum_{i=1}^L \sum_{j=1}^L \tr \left( \bfX_j^\HT \,\bfF_j\, \left(\bfF^{\,\HT}
  \, \bfF \right)^{-1} \, \bfF_i^{\,\HT}\,\bfX_i \right).
\label{eq:form2_expansion_1}
\end{align}
Define $\bfF_{Z\setminus i} = \left[ \bfF_1^{\,\HT} \; \cdots \;
  \bfF_{i-1}^{\,\HT} \; \bfF_{i+1}^{\,\HT} \;  \cdots \;
  \bfF_{L}^{\,\HT} \; \right]^{\,\HT}$ to be the composite matrix with
\mbox{channel $i$} removed. We can write an $i=j$ term in
(\ref{eq:form2_expansion_1}) as
\begin{align}
  &\frac{1\EndStrut{0.55ex}}{\StartStrut{0.65ex} M\tr(\SZZ)}
  \tr \left( \bfX_i^\HT \,\bfF_i\, \left(\bfF^{\,\HT}
  \, \bfF \right)^{-1} \, \bfF_i^{\,\HT}\,\bfX_i  \right)  
   =\frac{1\EndStrut{0.55ex}}{\StartStrut{0.65ex} M\tr(\SZZ)}
  \tr \left( \bfX_i^\HT \,\bfF_i\, \left(\bfF_i^{\,\HT}\,\bfF_i + 
  \bfF_{Z\setminus i}^{\,\HT}\,\bfF_{Z\setminus i} \right)^{-1} \,
  \bfF_i^{\,\HT}\, \bfX_i \right).
\end{align}
Now apply the matrix inversion lemma to write this as
\begin{align}
  \frac{1\EndStrut{0.55ex}}{\StartStrut{0.65ex} M\tr(\SZZ)}
 & \tr \left( \bfX_i^\HT \,\bfF_i\,\left(\bfF^{\,\HT}
  \, \bfF\right)^{-1} \, \bfF_i^{\,\HT}\,\bfX_i \right)  \nonumber \\
  &\quad =\frac{1\EndStrut{0.55ex}}{\StartStrut{0.65ex} M\tr(\SZZ)}
  \tr \bigg( \bfX_i^\HT \,\bfF_i\,\left[
    (\bfF_i^{\,\HT}\,\bfF_i)^{-1}
    -  (\bfF_i^{\,\HT}\,\bfF_i)^{-1}  \, \right. \left.\left( 
    (\bfF_i^{\,\HT}\,\bfF_i)^{-1}  + 
    (\bfF_{Z\setminus i}^{\,\HT}\,\bfF_{Z\setminus i})^{-1}
    \right)^{-1} 
    (\bfF_i^{\,\HT}\,\bfF_i)^{-1}\right]
  \bfF_i^{\,\HT}\, \bfX_i \bigg) \nonumber \\
  &\quad = \frac{1\EndStrut{0.55ex}}{\StartStrut{0.65ex}
    M\tr(\SZZ)} \tr \left(
  \bfX_i^{\,\HT} \, \bfP_{\bfF_i} \, \bfX_i \right) -
  [\,\bfN_{Z,\text{CFAR}}\,]_{ii} \nonumber \\[1.5ex]
  &\quad =
  \frac{\tr(\bfS_{ii})\EndStrut{0.55ex}}{\StartStrut{0.65ex} \tr(\SZZ)}
  \frac{\tr( \bfP_{\bfH_i} \, \bfS_{ii} )\EndStrut{0.55ex}}
       {\StartStrut{0.65ex}\tr(\bfS_{ii})}
       - [\,\bfN_{Z,\text{CFAR}}\,]_{ii} \nonumber \\[1.5ex]
 &\quad = \alpha_i \Lambda_{i,\text{CFAR}} -
  [\,\bfN_{Z,\text{CFAR}}\,]_{ii}.
\end{align}

\subsection{Decomposition of two-channel composite detectors: form 3 }
\label{sec:MultiChan_decomp_projection_form}

A second form for the cross validation matrix $\bfV_Z$ can be found by
writing the error matrix as
\begin{align}
  \bfE_Z &= \AXhat - \AYhat \nonumber \\
  &= \left[ \, (\bfF_X^{\,\HT} \, \bfF_X)^{-1} \,
    \bfF_X^{\,\HT},  \:\:
          -(\bfF_Y^{\,\HT} \, \bfF_Y)^{-1} \,
          \bfF_Y^{\,\HT} \, \right]\bfZ \nonumber \\
      &\eqdef \bfB_Z^{\,\HT} \, \bfZ
\end{align}
and noting that $\QEE = \bfB_Z^{\,\HT} \, \bfB_Z$.  Then
\begin{equation}
  \bfV_Z = \bfZ^{\,\HT}\, \bfP_{\bfB_Z} \, \bfZ.
\end{equation}
Then from Section~\ref{sec:MultiChan_decomp} we have
\begin{equation}
  \bfZ^{\,\HT} \, \bfP_{\bfF} \, \bfZ = 
  \bfX^{\,\HT} \, \bfP_{\bfF_X} \, \bfX + 
  \bfY^{\,\HT} \, \bfP_{\bfF_Y} \, \bfY - 
  \bfZ^{\,\HT} \, \bfP_{\bfB_Z} \, \bfZ.
\end{equation}
The matrix $\bfB_Z$ is such that $[\bfF_X^{\,\HT} \:\: \bfF_Y^{\,\HT} ]
\bfB_Z = \bfzero$.

\subsection{Decomposition of multi-channel detectors :
 assuming $\bfR_{\,\ell\ell}  = \sigma_\ell^2 \, \bfI, \:
 \bfH_\ell^{\,\HT}\,\bfH_\ell = \bfI_J$ }
\label{sec:unknown_gains_appendix}

Recall that the detector in (\ref{eq:cross_validation_term_1}) is
\begin{equation}
  \Lambda_{Z,\text{CFAR}} = \frac{\tr \left( \bfP_{\bfF} \, \SZZ
    \right)}{\tr(\SZZ)} 
\label{eq:Lambda_def2}
\end{equation}
and note that under the conditions of this section 
\begin{align}
  \bfF^{\,\HT}\, \bfF &=
  \sum_{\ell=1}^{L} \magsq{\, g_\ell}
  \bfH_\ell^{\,\HT} \, \bfH_\ell \nonumber \\
  &= 
  \sum_{\ell=1}^{L} \magsq{ \,g_\ell}\, \bfI_J \eqdef
  (\bfg^{\,\HT} \, \bfg\,) \, \bfI_J
\end{align}
where
\begin{equation}
  \bfg^{\,\HT} \eqdef [ \,
   g_1^{\,\mbox{\normalsize$\ast$}} \:
   g_2^{\,\mbox{\normalsize$\ast$}} \: \cdots \:
   g_L^{\,\mbox{\normalsize$\ast$}} \, ].
  \label{eq:gvec_def}
\end{equation}
The matched filter term contained in (\ref{eq:Lambda_def2}) is
\begin{equation}
  \bfF^{\,\HT}\, \bfZ = \sum_{\ell = 1}^L 
   g_\ell^{\,\mbox{\normalsize$\ast$}} \, \bfH_\ell^{\,\HT} \,
   \bfX_\ell.
\end{equation}
It follows that (\ref{eq:Lambda_def2}) can now be written as
\begin{align}
  &\Lambda_{Z,\text{CFAR}} \nonumber \\
  &\:=
  \frac{1}{ \bfg^{\,\HT} \, \bfg} \left[
   \frac{1}{M} \sum_{i=1}^L \sum_{j=1}^L 
   g_i^{\,\mbox{\normalsize$\ast$}} \, g_j \,
   \left( \frac{1}{\tr(\SZZ)} \,\tr(\bfH_i^{\,\HT} \,
   \bfX_i \, \bfX_j^\HT \, \bfH_j ) \right)  \right] \nonumber \\
  &\:\eqdef  \frac{ \bfg^{\,\HT} \, \bfM_{Z,\text{CFAR}} \, \bfg}
       {\bfg^{\,\HT} \, \bfg}.
\label{eq:general_Rayleigh_quotient_cfar}
\end{align}
A similar Rayleigh quotient form, for a special case of this, was presented in
\cite{VanKay12}.
Define the coherence term 
\begin{align}
 c_{ij} &\eqdef    \frac{\tr \left( \bfH_i^{\,\HT}\, \bfX_i \,
   \bfX_j^{\,\HT}\, \bfH_j \right) }
    { \left( \tr \left( \bfH_{i\phantom{j}}^{\,\HT}\, \bfX_i \, \bfX_i^{\,\HT} \,
      \bfH_i \right) \right)^{1/2} \,
      \left( \tr \left( \bfH_j^{\,\HT}\, \bfX_j \, \bfX_j^{\,\HT} \, \bfH_j
      \right) \right)^{1/2}}.
    \label{eq:coherence_def}
\end{align}
Note that this term is invariant to \emph{different} scalings
of the data $\bfX_i$ and $\bfX_j$.  We can write an element of
$\bfM_{Z,\text{CFAR}}$ in (\ref{eq:general_Rayleigh_quotient_cfar}) as
\begin{align}
 \left[ \bfM_{Z,\text{CFAR}}\right]_{ij} &= 
  \frac{1}{\tr(\SZZ)} \,
   \left( \frac{1}{M} \tr (\bfH_i^{\,\HT}\, \bfX_i \, \bfX_i^{\,\HT} \,
   \bfH_i) \right)^{1/2} \,    \left( \frac{1}{M} \tr (\bfH_j^{\,\HT}\, \bfX_j \, \bfX_j^{\,\HT} \,
   \bfH_j) \right)^{1/2} \! c_{ij} \nonumber \\[1.5ex]
   &=
   \frac{1}{\tr(\SZZ)} \,
   \left( \tr (\bfP_{\bfF_i}\, \bfS_{ii}) \right)^{1/2} \,
   \left( \tr (\bfP_{\bfF_j}\, \bfS_{jj}) \right)^{1/2} \!
   c_{ij} \nonumber \\[1.5ex]
  &=
   \left(\frac{\tr(\bfS_{ii})}{\tr(\SZZ)}\right)^{1/2} \,
   \left(\frac{\tr(\bfS_{jj})}{\tr(\SZZ)}\right)^{1/2} \, \left( \frac{\tr (\bfP_{\bfH_i}
     \,\bfS_{ii})}{\tr(\bfS_{ii})} \right)^{1/2} \,
   \left( \frac{\tr (\bfP_{\bfH_j} \bfS_{jj})}{\tr(\bfS_{jj})} \right)^{1/2} \!
   c_{ij} \nonumber \\[1.8ex]
  &= \alpha_i^{1/2} \, \alpha_j^{1/2} \, \Lambda^{1/2}_{i,\text{CFAR}} \, 
   \Lambda^{1/2}_{j,\text{CFAR}}\, c_{ij}.
\label{eq:Mij}
\end{align}
This a general quadratic form under the conditions
of this section and it applies to those detectors in
Section~\ref{sec:Unknown_channel_gains}.  The definitions of 
$\alpha_i$  and $\Lambda_i$ change for different noise models.

We can express the detector in (\ref{eq:general_Rayleigh_quotient_cfar}) 
in our canonical form by noting
\begin{align}
\frac{\bfg^{\,\HT} \, \bfM_{Z,\text{CFAR}} \,\bfg}{\bfg^{\,\HT}\, \bfg} &=
\frac{\bfg^{\,\HT} \, \left[\, \diag\{ 
    \,\bfM_{Z,\text{CFAR}}\}
  \,\right] \,\bfg}{\bfg^{\,\HT}\, \bfg}   +
\frac{\bfg^{\,\HT} \, \left[\, \bfM_{Z,\text{CFAR}}-\diag\{ 
    \bfM_{Z,\text{CFAR}}\}\, \right]
  \,\bfg}{\bfg^{\,\HT}\, \bfg}  \nonumber \\
&\eqdef \sum_{\ell=1}^L \frac{ \magsq{\,g_\ell\,}}{\bfg^{\,\HT}\, \bfg} \alpha_\ell
\Lambda_{\ell,\text{CFAR}}\; +\;
\frac{\bfg^{\,\HT} \, \bfM_2 \, \bfg}{\bfg^{\,\HT} \, \bfg}
  \label{eq:pre_canonical_expansion1}
\end{align}
where $\bfM_2$ is identical to $\bfM_{Z,\text{CFAR}}$ except for
the diagonal terms, which are zero in the former.  Now note
\begin{equation}
  \frac{\magsq{\,g_\ell\,}}{\bfg^{\,\HT} \, \bfg} \alpha_\ell
  \Lambda_{\ell,\text{CFAR}}
  = \alpha_\ell \Lambda_{\ell,\text{CFAR}} - \sum_{i \neq \ell}^L
  \frac{\magsq{\,g_i\,}}{\bfg^{\,\HT} \, \bfg},
\end{equation}
which allows (\ref{eq:pre_canonical_expansion1}) to be written as
\begin{align}
\Lambda_{Z,\text{CFAR}}&=\frac{\bfg^{\,\HT} \, \bfM_{Z,\text{CFAR}} \,\bfg}{\bfg^{\,\HT}\, \bfg} =
\sum_{\ell=1}^L \alpha_\ell \Lambda_{X_\ell,\text{CFAR}} - \frac{ \bfg^\HT \,
  \bfT_{Z,\text{CFAR}} \, \bfg}{\bfg^{\,\HT} \, \bfg} \nonumber \\
&= \sum_{\ell=1}^L \alpha_\ell \Lambda_{X_\ell,\text{CFAR}} - V_{Z,\text{CFAR}},
\end{align}
which is our canonical form for the conditions of this section.
The elements of $\bfT_{Z,\text{CFAR}}$ are
\begin{align}   
 [\,\bfT_{Z,\text{CFAR}}\,]_{\,ii}   &=
  \sum_{\ell \neq i}^L \alpha_\ell \Lambda_{X_\ell,\text{CFAR}}; \nonumber \\
  [\,\bfT_{Z,\text{CFAR}}\,]_{\,ij}  &=
      - [\,\bfM_{Z,\text{CFAR}}\,]_{\,ij}  \quad (i \neq j) \nonumber
      \\[1.3ex]
  &= -\alpha_i^{1/2} \, \alpha_j^{1/2}\,
      \Lambda_{i,\text{CFAR}}^{1/2}\,
      \Lambda_{j,\text{CFAR}}^{1/2}\,   \, c_{ij}.
\end{align}  

\bibliographystyle{IEEEtran}
\bibliography{TwoChannelRefs}

\end{document}

%% file: basedef2e.tex
\newcommand{\bfa}{{\mathbf{a}}}
\newcommand{\bfA}{{\mathbf{A}}}

\newcommand{\bfB}{{\mathbf{B}}}

\newcommand{\bfD}{{\mathbf{D}}}

\newcommand{\bfE}{{\mathbf{E}}}

\newcommand{\bfF}{{\mathbf{F}}}
\newcommand{\bfg}{{\mathbf{g}}}

\newcommand{\bfH}{{\mathbf{H}}}

\newcommand{\bfI}{{\mathbf{I}}}

\newcommand{\bfK}{{\mathbf{K}}}

\newcommand{\bfM}{{\mathbf{M}}}

\newcommand{\bfN}{{\mathbf{N}}}

\newcommand{\bfP}{{\mathbf{P}}}

\newcommand{\bfQ}{{\mathbf{Q}}}

\newcommand{\bfR}{{\mathbf{R}}}

\newcommand{\bfS}{{\mathbf{S}}}

\newcommand{\bfT}{{\mathbf{T}}}
\newcommand{\bfu}{{\mathbf{u}}}
\newcommand{\bfU}{{\mathbf{U}}}

\newcommand{\bfV}{{\mathbf{V}}}

\newcommand{\bfx}{{\mathbf{x}}}
\newcommand{\bfX}{{\mathbf{X}}}

\newcommand{\bfY}{{\mathbf{Y}}}
\newcommand{\bfz}{{\mathbf{z}}}
\newcommand{\bfZ}{{\mathbf{Z}}}
\newcommand{\bfzero}{{\boldsymbol{0}}}

\newcommand{\bfLambda}{{\boldsymbol{\Lambda}}}

\newcommand{\bfSigma}{{\boldsymbol{\Sigma}}}

\newcommand{\bfPhi}{{\boldsymbol{\Phi}}}

\renewcommand{\hbar}{{\bar{h}}}

\newcommand{\sigmawhat}{{\widehat{\sigma}}}

\newcommand{\bfAwhat}{{\widehat{\mathbf{A}}}}

\newcommand{\bfFwhat}{{\widehat{\mathbf{F}}}}

\newcommand{\bfHwhat}{{\widehat{\mathbf{H}}}}

\newcommand{\Awtil}{{\widetilde{A}}}

\newcommand{\gwtil}{{\widetilde{g}}}
\newcommand{\Gwtil}{{\widetilde{G}}}

\newcommand{\Vwtil}{{\widetilde{V}}}

\newcommand{\Deltawtil}{{\widetilde{\Delta}}}

\newcommand{\Lambdawtil}{{\widetilde{\Lambda}}}

\newcommand{\bfFwtil}{{\widetilde{\mathbf{F}}}}
\newcommand{\bfgwtil}{{\widetilde{\mathbf{g}}}}

\newcommand{\bfMwtil}{{\widetilde{\mathbf{M}}}}

\newcommand{\bfQwtil}{{\widetilde{\mathbf{Q}}}}

\newcommand{\bfSwtil}{{\widetilde{\mathbf{S}}}}

\newcommand{\bfTwtil}{{\widetilde{\mathbf{T}}}}

\newcommand{\bfUwtil}{{\widetilde{\mathbf{U}}}}

\newcommand{\bfxwtil}{{\widetilde{\mathbf{x}}}}
\newcommand{\bfXwtil}{{\widetilde{\mathbf{X}}}}

\newcommand{\bfzwtil}{{\widetilde{\mathbf{z}}}}
\newcommand{\bfZwtil}{{\widetilde{\mathbf{Z}}}}

\newcommand{\calC}{{\mathcal{C}}}

\newcommand{\calE}{{\mathcal{E}}}

\newcommand{\calL}{{\mathcal{L}}}

\newcommand{\calN}{{\mathcal{N}}}

\newcommand{\C}{{\mathbb{C}}}

\newcommand{\abs}[1]{\lvert #1 \rvert}
\newcommand{\magsq}[1]{\lvert #1 \rvert^{\,2}}

\newcommand{\tr}{{\text{tr}}}

\newcommand{\diag}{{\text{diag}}}

\newcommand{\subspace}[1]{\langle #1 \rangle}

\newcommand{\eqdef}{{\ \overset{\triangle}{=}\ }}

\newcommand{\addots}{\mathinner{\mkern1mu\raise1pt\vbox{\kern7pt\hbox{.}}
   \mkern2mu \raise4pt\hbox{.}\mkern2mu\raise7pt\hbox{.}\mkern1mu}}

\newcommand{\HT}{{\mathrm{H}}}

\newcommand{\twovec}[2]{\left[ \begin{array}{c} #1 \\ #2 \end{array} \right] }

\newcommand{\fourvec}[4]{\left[ \begin{array}{c} #1 \\ #2 \\ #3 \\ #4 \end{array} \right] }

\newcommand{\twovecstretch}[3]{\left[ {\renewcommand{\arraystretch}{#3}\begin{array}{c} #1 \\ #2 \end{array}} \right] }

\newcommand{\twomatstretch}[5]{\left[ {\renewcommand{\arraystretch}{#5}\begin{array}{cc} #1 & #2 \\ #3 & #4 \end{array}} \right] }

